\documentclass[prc,aps,floats,floatfix,twocolumn,showpacs,nofootinbib,%
superscriptaddress]{revtex4}

\usepackage{amsmath}
\usepackage{amssymb}
\usepackage{amsfonts}
\usepackage{booktabs}
\usepackage{graphicx}
\usepackage{tabularx}
\usepackage{multirow}
\usepackage{lscape}
\usepackage{rotating}

\usepackage{pstricks}

\begin{document}

\title{Spurious finite-size instabilities in nuclear energy density functionals: spin channel}


\author{A. Pastore}
\affiliation{CEA, DAM, DIF, F-91297 Arpajon, France}

\author{D. Tarpanov}
\affiliation{Institute of Theoretical Physics, Faculty of Physics,University of Warsaw, ul. Pasteura 5,
02-093 Warsaw, Poland}
\affiliation{Institute for Nuclear Research and Nuclear Energy, Bulgarian Academy of Sciences, BG-1784 Sofia, Bulgaria}

\author{D. Davesne}
\affiliation{Universit\'e de Lyon, F-69003 Lyon, France; Universit\'e Lyon 1,
             43 Bd. du 11 Novembre 1918, F-69622 Villeurbanne Cedex, France   \\
             CNRS-IN2P3, UMR 5822, Institut de Physique Nucl{\'e}aire de Lyon}
             
\author{J. Navarro}
\affiliation{IFIC (CSIC-Universidad de Valencia), Apdo. 22085, E-46.071-Valencia, Spain}

\begin{abstract}
\begin{description}
\item[Background] It has been recently shown, that some Skyrme functionals can lead to non-converging results in the calculation of some properties of atomic nuclei. A previous study has pointed out a possible link between these convergence problems and the appearance  of finite-size instabilities in symmetric nuclear matter (SNM) around saturation density. 
\item[Purpose] We show that the finite-size instabilities not only affect the ground state properties of atomic nuclei, but they can also influence the calculations of vibrational excited states in finite nuclei.
\item[Method] We perform systematic fully-self consistent Random Phase Approximation (RPA) calculations in spherical doubly-magic nuclei. We employ several Skyrme functionals and vary the isoscalar and isovector coupling constants of the time-odd term $\mathbf{s}\cdot \Delta \mathbf{s}$ . We determine critical values of these coupling constants beyond which the RPA calculations do not converge because RPA the stability matrix becomes non-positive.
\item[Results] By comparing the RPA calculations of atomic nuclei with those performed for SNM we establish a correspondence between the critical densities in the infinite system and the critical coupling constants for which the RPA calculations do not converge.
\item[Conclusions] We  find a quantitative stability criterion to detect finite-size instabilities related to the spin $\mathbf{s}\cdot \Delta \mathbf{s}$ term of a functional. This criterion could be easily implemented into the standard fitting protocols to fix the coupling constants of the Skyrme functional.
\end{description}
\end{abstract}


 
\date{\today}


\maketitle

%
\section{Introduction}
\label{sect:intro}

The Nuclear Energy Density Functional (NEDF) is a standard tool used to describe properties of atomic nuclei from drip-line to drip-line~\cite{ben03,erl12}. Among the different functionals available nowadays, the one derived from the non-relativistic Skyrme interaction is one of the most popular.
A very important aspect for the determination of a functional is the optimization procedure used to obtain its effective coupling constants~\cite{kor10,kor12,kor13}.
To this purpose one has to minimize in a multi-dimensional space a penalty function. The latter is obtained by considering some \emph{observables} of finite nuclei as for example masses and radii and some \emph{pseudo-observables} of infinite nuclear matter~\cite{chab97,was12,cha15}.
The minimization procedure can be quite evolved since the penalty function presents discontinuities in the parameter space and therefore may induce some difficulties in finding a stable minimum.
For a more detailed discussion, we refer the reader to Ref.~\cite{dob14} and references therein.

Since the input parameters may present some uncertainties or their number is limited, a coupling constant can be poorly constrained making the extrapolation procedure very problematic~\cite{erl10}. Such a situation can also eventually  lead to unphysical instabilities. By instance, in Ref.~\cite{mar02}, Margueron \emph{et al.} have shown that most of the Skyrme functionals present  spin or ferromagnetic instabilities, that is a spontaneous and complete polarization of the infinite medium. The latter is not predicted by any other model either phenomenological or \emph{ab-initio}~\cite{nav13}, thus leading to the conclusion that it is a pathology of the Skyrme functional itself.
These type of instabilities can be simply controlled by inspecting the properties of the corresponding Landau parameters derived from the Skyrme functional~\cite{dav14F} during the optimization procedure~\cite{cha10}.
However, another type of instability, not related to the entire system, but specific to finite-size domains  with typical size $\lambda\approx {2\pi} / {q}$, where $q$ is the transferred momentum has been recently detected by Lesinski \emph{et al.}~\cite{les06} for the SKP~\cite{dob84} and LNS~\cite{cao06} functionals. Contrary to the previous case, since the instability occurs at non-zero values of $q$, it can not be detected by exploring the Landau limit. 

In Ref.~\cite{hel13}, Hellemans \emph{et al.}  investigated the nature of finite-size instabilities for the scalar-isovector channel. The authors have established a simple relation between the presence of these instabilities in finite nuclei calculations and the position of zero-energy modes  calculated using  Random-Phase-Approximation in infinite nuclear matter~\cite{pas14D}.
In particular, it has been shown that whenever the critical density $\rho_{c}$ at which a pole ($i.e$ zero energy mode with infinite strength) appears in SNM is close to the saturation density of the system, that instability manifests itself in finite nuclei leading to non-physical solutions. For example, in the scalar-isoscalar case it  leads to a complete separation between protons and neutrons.
Such a relation is particularly important since RPA calculations in SNM~\cite{pas14D} are analytical and not time consuming. Recently, they implemented a criterion based on RPA in SNM into a fitting procedure to prevent, by construction, this kind of problems~\cite{pas13}.
 
In the present article, we explore deeper the question of the correspondence between poles in  the infinite medium and instabilities in finite nuclei. In particular, we want to know whether this remains valid in the vector channel of the interaction.
Our study is motivated by the problems detected recently using Time-Dependent-Hartree-Fock methods (TDHF)~\cite{fra12}  and Cranked-Hartree-Fock-Bogoliubov (CHFB) calculations~\cite{hel12}, which clearly exhibit problems of convergence when using some Skyrme functionals.
In particular, in Ref.~\cite{hel12}, the instability problem has been clearly associated with the time odd terms which depend on the laplacian and gradient of the spin density~\cite{per04}.
In the present article, we follow the protocol outlined in Ref.~\cite{hel13}, but using the new spherical RPA code recently developed by the FIDIPRO group~\cite{toi10,car12}. The main advantages of using RPA are that the time-odd terms~\cite{dob95} are active and that it is numerically less time-consuming and thus more adapted for systematic calculations
as compared to other methods such as CHFB or TDHF.

The article is organized as follows: in Sec.\ref{sec:skyrme} we briefly sketch the formalism of the Skyrme functionals, in Sec.\ref{sec:rpa} we present the method used to detect and identify the appearance of finite-size instabilities in finite nuclei, while in Sec.\ref{rpa:snm} we present the formalism of RPA in infinite matter. The link between the two methods is presented in Sec.\ref{sec:nuclei}, where a simple criterion to detect instabilities is also proposed. Finally we present our conclusions in Sec.\ref{sec:conclusion}.

\section{Skyrme functionals}\label{sec:skyrme}

The Skyrme functional, $\mathcal{E}_{Sk}$, is a linear combination of coupling constants and local densities of the form

\begin{eqnarray}
\mathcal{E}_{Sk}&=&\int d^{3}r \sum_{t=0,1}\left\{ C^{\rho}_{t}[\rho_{0}]\rho_{t}^{2}+C^{s}_{t}[\rho_{0}]\mathbf{s}^{2}_{t}+C^{\Delta\rho}_{t}\rho_{t}\Delta\rho_{t}\right.\nonumber\\
&+&C^{\nabla s}_{t}(\nabla\cdot \mathbf{s}_{t})^{2}+C^{\Delta s}_{t}\mathbf{s}_{t}\Delta \mathbf{s}_{t}+C^{\tau}_{t}(\rho_{t}\tau_{t}-\mathbf{j}_{t}^{2})\nonumber\\
&+&C^{T}_{t}\left( \mathbf{s}_{t}\cdot \mathbf{T}_{t}-\sum_{\mu,\nu=x}^{z}J_{t,\mu\nu}J_{t,\mu\nu}\right)\nonumber\\
&+&C^{F}_{t}\left[ \mathbf{s}_{t}\cdot \mathbf{F}_{t}-\frac{1}{2}\left(\sum_{\mu,\nu=x}^{z}J_{t,\mu\nu}\right)^{2}-\frac{1}{2}\sum_{\mu,\nu=x}^{z}J_{t,\mu\nu}J_{t,\nu\mu}\right]\nonumber\\
&+&\left. C^{\nabla J}_{t}(\rho_{t}\nabla \mathbf{J}_{t}+\mathbf{s}_{t}\nabla \times \mathbf{j}_{t} )\right\}\;.
\end{eqnarray}

The definition of the local densities $\rho_{t},\mathbf{s}_{t},...$ is standard and can be found in Refs.\cite{per04,les07}. 
The density dependent coupling constants are written as

\begin{eqnarray}
C^{\rho}_{t}[\rho_{0}]&=&C^{\rho,0}_{t}+\rho_0^{\alpha}C^{\rho,\alpha}_{t}\,,\\
C^{s}_{t}[\rho_{0}]&=&C^{s,0}_{t}+\rho_0^{\alpha}C^{s,\alpha}_{t}\,.
\end{eqnarray}

The finite-size instabilities are mainly related to the presence of derivative terms: $\rho_{t}\Delta\rho_{t}$ for the scalar channel and  $\mathbf{s}_{t}\Delta \mathbf{s}_{t}$ for the vector one. The term $C^{\nabla s}_{t}(\nabla\cdot \mathbf{s}_{t})^{2}$ represents  also a possible source of instability as already mentioned in Ref.~\cite{hel12}, but it is active only  if the interaction contains an explicit tensor term. In the present analysis, we do not consider explicitly the instabilities related to such a term and we  focus more on the ones related to the time-odd  Laplacian term.

To perform our study,  we chose six representative Skyrme functionals: T44~\cite{les07}, SAMi~\cite{roc12}, BSk27~\cite{gor13bsk}, SIII~\cite{bei75}, SKO$'$~\cite{rei99} and SLy5~\cite{chab97}.
These functionals have been obtained by different groups using different penalty functions and they are currently used within the nuclear structure community to perform calculations of nuclear observables for both ground and excited states of nuclei along the nuclear chart.

\begin{figure}[!h]
\begin{center}
\includegraphics[width=0.40\textwidth,angle=-90]{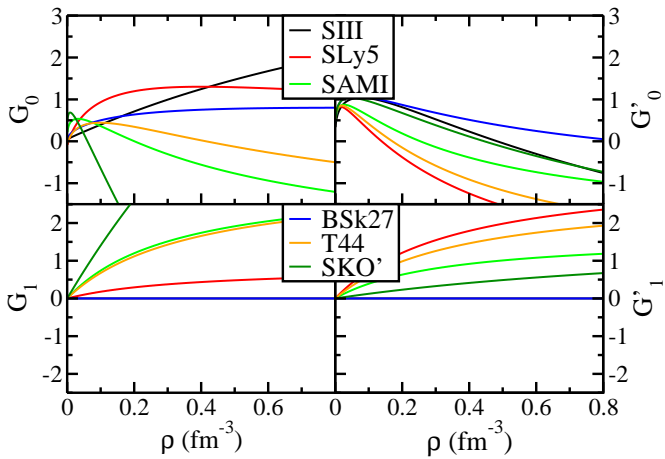}
\end{center}
\caption{(Color online) Evolution of Landau parameters $G_{l}$ and $G'_{l}$ $(l=0,1)$ as a function of the density in SNM for the six Skyrme functionals considered in the article.  }
\label{landauG}
\end{figure}

The presence of long-wavelength ferromagnetic instabilities in SNM can be easily determined by inspecting the behavior of the Landau parameters $G_{\ell=0,1}$ and $G^{'}_{\ell=0,1}$, which can be expressed in terms of Skyrme coupling constants as~\cite{pas12}
\begin{eqnarray}
N_{0}^{-1}G_{0}&=&2C^{s,0}_{0}+(2+\alpha)(1+\alpha)C^{\rho,\alpha}_{0}\rho_0^{\alpha}\nonumber \\
                          &+&2k_{F}^{2}C^{T}_{0}+ \frac{2}{3}k_{F}^{2}C^{F}_{0}\label{land1} \,,\\
N_{0}^{-1}G_{1}&=&-2k_{F}^{2}C^{T}_{0}-\frac{2}{3}k_{F}^{2}C^{F}_{0}\label{land2} \,,\\
N_{0}^{-1}G_{0}^{'}&=&2C^{s,0}_{1}+(2+\alpha)(1+\alpha)C^{\rho,\alpha}_{1}\rho_0^{\alpha}\nonumber \\
                            &+&2k_{F}^{2}C^{T}_{1}+\frac{2}{3}k_{F}^{2}C^{F}_{1}\label{land3} \,,\\
N_{0}^{-1}G_{1}^{'}&=&-2k_{F}^{2}C^{T}_{1}-\frac{2}{3}k_{F}^{2}C^{F}_{1}\label{land4} \,,
\end{eqnarray}
\noindent where $k_{F}$ the Fermi momentum and $N_{0}^{-1}=\frac{\hbar^{2}\pi^{2}}{2m^{*}k_{F}}$ is the density of states at the Fermi surface. In Fig.\ref{landauG}, we show the evolution of the Landau parameters $G_{l}$ and $G'_{l}$ in SNM for the six selected functionals.

According to the results of Ref.~\cite{mar02}, whenever the inequality
\begin{eqnarray}\label{ineq}
G_{l}>-(2l+1)\;,
\end{eqnarray}
\noindent is not satisfied, a static deformation of the Fermi surface against spontaneous polarization is observed.
In presence of an explicit tensor term new stability condition are necessary (see Refs.~\cite{bac79,nav13} for a more detailed discussion).
From Fig.~\ref{landauG}, we observe that the SKO$'$ functional is spin-unstable in the isoscalar channel at very low densities $\rho_{}\approx0.14$ fm$^{-3}$.
Such a problem manifests in RPA calculations of finite nuclei. We have selected the spherical doubly-magic nuclei $^{40}$Ca, $^{56}$Ni, $^{132}$Sn, and $^{208}$Pb as specific examples. In particular we have not been able to perform any complete calculation due to the appearance of several imaginary phonons which spoil the results. It is actually  possible to obtain a result only by reducing the available $ph$-space, but in that case the results strongly depend on the parameter space and on the related cut-off.
The use of small parameter space is the reason why other groups have been able to present results using this functional, although the physical value of such results is very doubtful. For such a reason the  SKO$'$ should not be used to perform RPA calculations.

The functionals SLy5 and T44 are also unstable, in the spin-isovector channel, but at much larger values of the density  which are not relevant for finite nuclei, although they can be important for astrophysical applications~\cite{cha09}.

It is worth noticing that the functionals SAMi and BSK27 have been built to remove these instabilities directly during their fitting procedure. Therefore, they respect Eq.~(\ref{ineq}) up to several times saturation density.

The functionals SIII and BSk27 give Landau parameters with $l=1$ identically equal to zero. This is due to the dropping of the so-called $J^{2}$ term in these functionals, $i.e.$ we have $C^T_{t=0,1}=0$. 
By inspecting Eqs.~(\ref{land1}-\ref{land4}), we observe that Landau parameters do not depend on the coupling constants $C^{\nabla s}_{t}$ and $C^{\Delta s}_{t}$.  
As discussed in Ref.~\cite{pas12}, these terms are proportional  to $q^2$ in the residual interaction, and thus they are zero in the limit  $q\rightarrow0$.

\section{RPA instabilities in finite nuclei}\label{sec:rpa}

We now focus our attention on the $\mathbf{s}_{t}\Delta \mathbf{s}_{t}$ terms, to analyze its effects on possible instabilities. The value of the $C^{\Delta s}_{t}$ coupling constants is given in Tab.~\ref{tab:skyrme} for the functionals considered in this article.

\begin{table}
\begin{center}
\caption{Values of the coupling constants $C^{\Delta s}_{t=0,1}$ and SNM saturation density $\rho_{sat}$ of the functionals considered here.}
\begin{tabular}{c|ccc}
\hline
\hline
         & $\rho_{\text{sat}} (\text{fm}^{-3})$ & $C^{\Delta s}_{0} (\text{MeVfm}^{5})$ & $C^{\Delta s}_{1} (\text{MeVfm}^{5})$\\[1mm]
\hline
  SIIII &0.145 &17.031 &17.031\\[1mm]
  SLy5&0.160 & 46.087&14.062\\[1mm]
  T44 &0.160 & 72.670&1.393 \\[1mm]
  BSk27 &0.158 &27.925 &16.691\\[1mm]
  SAMi &0.159 &44.720&20.962\\[1mm]
  SKO$'$& 0.160 &42.790 &16.553\\[1mm]
  \hline 
  \hline
\end{tabular}
\label{tab:skyrme}
\end{center}
\end{table}

To detect the presence of finite-size instabilities, we perform RPA calculations in the selected set of nuclei  using the numerical code described in Ref.~\cite{toi10} .
The calculations are fully self-consistent, $i.e.$ include both all the terms of the functional and all the states within the basis. In the present article we consider a basis of 16 major oscillator shells. No further cut-offs are included in the calculations as commonly done, see for example Refs.~\cite{ter05,los10,lef15} for discussion.
It is important to stress that although the use of a cut-off in $ph$ space can be safely adopted for \emph{stable} calculations, this could hide the problems discussed here.
To observe the presence of instabilities, we calculate the stability matrix $\mathcal{S}$~\cite{Book:Ring1980}

\begin{eqnarray}
\mathcal{S}=\left( \begin{array}{cc}
A & B^{*}\\
B^{*} & A^{*}
\end{array}\right)\;,
\end{eqnarray}

\noindent which should be positive-defined in the case of RPA. Negative or imaginary eigenvalues of the $\mathcal{S}$ matrix are thus the signal of finite-size instabilities, which are related to the fact that the ground state configuration we have used to calculate the vibrational band is not a minimum of the energy surface. As a consequence, the system is no more stable against small fluctuations around this point.
To study the impact of the term $\mathbf{s}_{0}\Delta \mathbf{s}_{0}$, we put to zero the coupling constant $C^{\Delta s}_{1}$ and we vary  $C^{\Delta s}_{0}$ and vice versa through a multiplicative parameter $\gamma$.
Similarly to what has been done in Ref.~\cite{hel13} for the term $\rho_{1}\Delta\rho_{1}$, we want to determine the critical value of the coupling constant beyond which imaginary eigenvalues appear.

\begin{figure}[!h]
\begin{center}
\includegraphics[width=0.5\textwidth,angle=0]{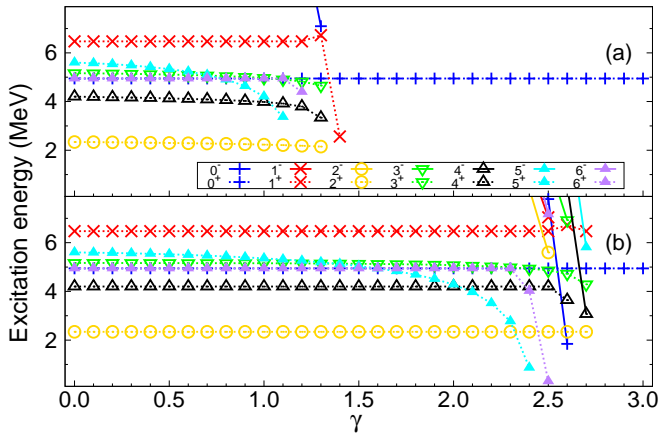}
\end{center}
\caption{(Color online)  Evolution of the lowest energy vibrational states for different multipolarities $J^{\pi}$ in $^{56}$Ni as a function of the isoscalar coupling constant  $C^{\Delta s}_{0}$ (panel a) and the isovector one (panel b), for for the SAMi functional. }
\label{phononsISOsami}
\end{figure}

In Fig.~\ref{phononsISOsami} (a), we show the evolution of the lowest phonon energies in $^{56}$Ni as a function of the parameter $\gamma$ which multiplies the isoscalar coupling constant for the SAMi functional. For $\gamma=1$ we obtain the nominal value of the coupling constant as shown in Tab.~\ref{tab:skyrme}.
When we modify $\gamma$, we are effectively building a series of different functionals which differ only by the value of this time-odd coupling constant, and for each functional we solve the corresponding RPA equations, obtaining the energies of excitations for multipolarities up to $J = 6$ and both parities $\pi=\pm1$. 
All these newly constructed functionals have \emph{exactly} the same bulk properties of infinite matter, ground state properties of even-even nuclei and Landau parameters.

\noindent When the multiplicator is in the range $\gamma \in[0:0.8]$, the phonon spectrum is essentially flat, telling us that the phonons are not so sensitive to the exact value of this coupling constant, but when we approach $\gamma\approx1$ ($i.e.$ the nominal value), we notice a sudden bending in the $J=5^{+}$ phonon.
Going beyond the nominal value, we observe that there is a change in the ordering of the lowest excited states until we reach a critical value at which the RPA calculations are no more trustful since some of the eigenvalues become imaginary.
The critical value is not the same for all different multipolarities, still such a value exists for all multipolarities and we observe a concentration of the strength of the low-lying RPA phonons  when the respective critical value is approached.
We remind that the breathing mode $0^{+}$ does not break the time-reversal symmetry thus the time-odd coupling constants play no-role~\cite{ves12}.

We have also analyzed the behavior of the vibrational states up to $J = 12$ and observed the same trend. It also seems that they are not decisive for setting the critical value of the $C^{\Delta s}_{t}$ coupling constant.
For this particular parameterization, the isoscalar coupling constant is closer to its critical value for most of the multipolarities considered here.

In Fig.~\ref{phononsISOsami} (b), we repeat the same calculations, but for the isovector coupling constant  $C^{\Delta s}_{1}$. The multiplier parameter $\gamma$ is set in exactly the same way, but for the isovector constant. In this case the critical coupling constant is located at $\approx2.5$ its nominal value.

\begin{figure*}[!h]
\begin{center}
\includegraphics[width=0.9\textwidth,angle=0]{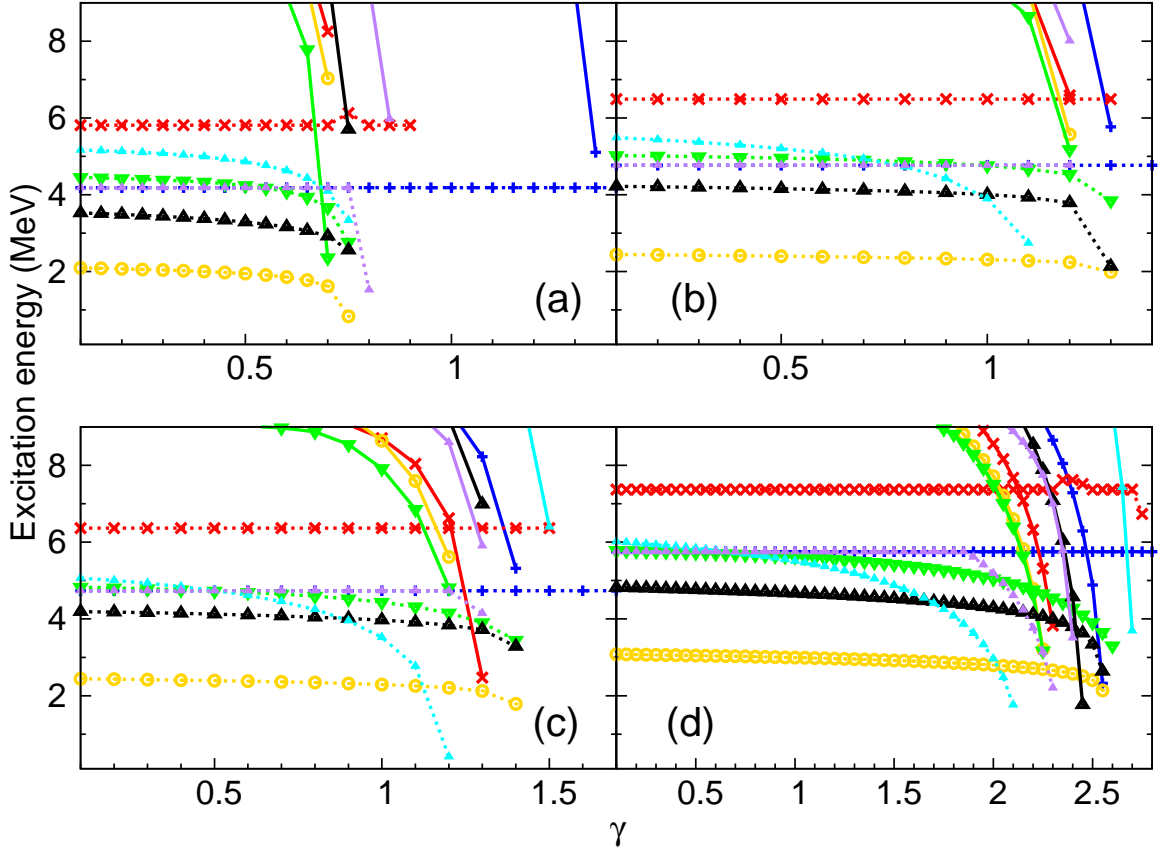}
\end{center}
\caption{(Color online). Evolution phonons in $^{56}$Ni as a function of the multiplicative  factor $\gamma$ for T44 (panel a), SLy5 (panel b), BSk27 (panel c) and SIII (panel d). The legend is the  the same as for Fig.\ref{phononsISOsami}.}
\label{phononsISO2}
\end{figure*}

In Fig.~\ref{phononsISO2}, we investigate further the instabilities related to the isoscalar coupling constant in $^{56}$Ni, by performing systematic RPA calculations with the other selected functionals.
We observe that T44 is not stable when we use the nominal value of the coupling constant, making it improper for RPA calculations. To converge the RPA calculations, we have to use a smaller multiplier $\gamma \leq 0.5$.
Since in the present study, we do not modify the coupling constants related to the term  $(\nabla s_{t=0,1})^{2}$, our result can not be directly compared with the one obtained in Ref.~\cite{hel13}.

Similarly the isoscalar coupling constants of SLy5 and BSk27 parameterizations of the Skyrme force are close to the limit of stability and we clearly observe that the phonons in the area where $\gamma \approx 1$ strongly depend on the exact value of the time-odd coupling constant $C^{\Delta s}_{0}$.
We can thus conclude that these functionals are not adapted to describe vibrational states in finite nuclei.

\begin{figure}
\begin{center}
\includegraphics[width=0.42\textwidth,angle=-0]{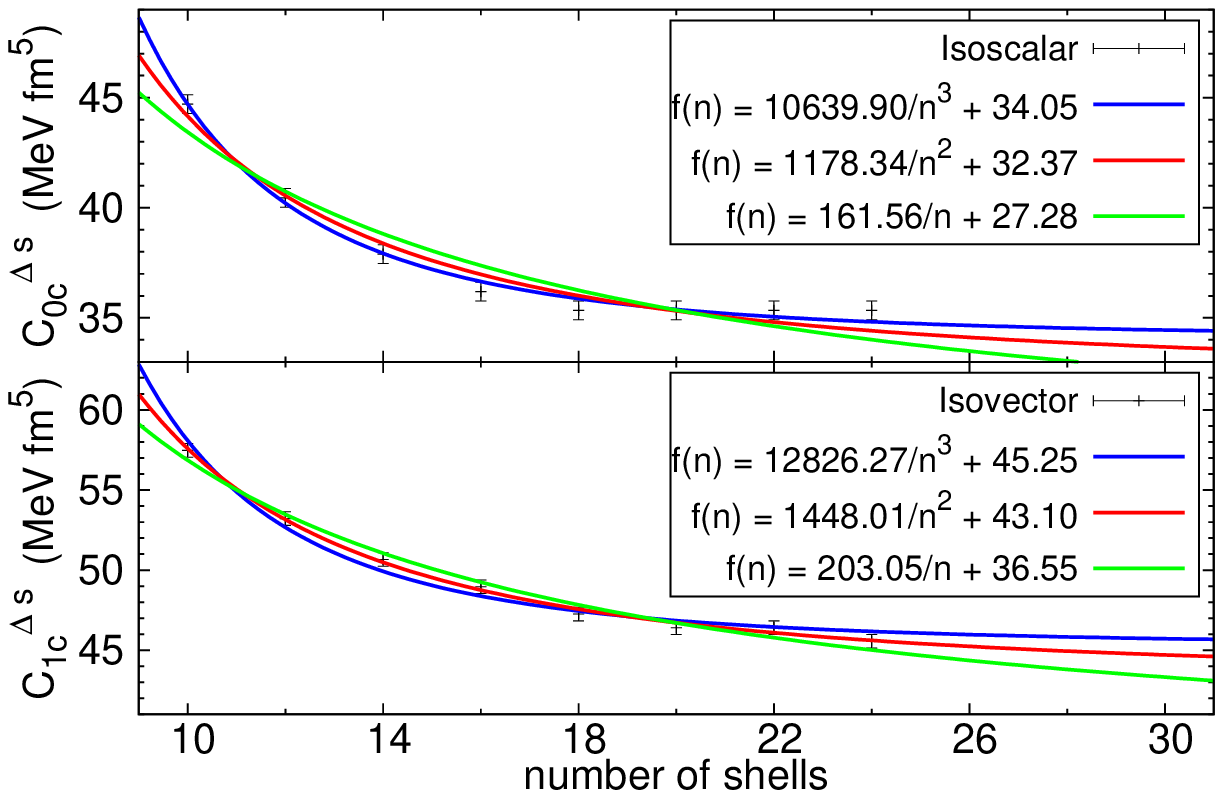}
\end{center}
\caption{(Color online) Evolution of the critical coupling constants for the SIII functional for different number of shells included in the calculation. With lines we have shown some proposed functions to describe the dependence of the critical coupling constants with the shell numbers.}
\label{shell}
\end{figure}

We have also tested the dependence of our results on the size of the basis.
In Fig.~\ref{shell}, we show the evolution of the lowest critical coupling constant $C^{\Delta s}_{0c}$ (panel a) and $C^{\Delta s}_{1c}$ (panel b), $i.e.$ the value beyond which at least one of the multipolarities studied here gives an imaginary phonon, as a function of the number of major shells $n$ included in our calculation for the SIII functional in $^{56}$Ni. Since we perform the calculations  on a discrete mesh of the multiplicative factor $\gamma$, we have defined the point of collapse as the value in between the last point $\gamma_{N-1} $ for which all the phonon energies are real and the first point  $\gamma_{N} $ at which we have a collapsed solution. We thus define $\gamma_c =(\gamma_{N} + \gamma_{N-1}) / 2$ with the error bars set by the values of $\gamma_{N}$ and  $\gamma_{N-1}$.

We observe a strong dependence of the coupling constant  $C^{\Delta s}_{0c}$ on the number of shells used in the calculations: the larger $n$, the smaller $C^{\Delta s}_{0c}$. Such a feature has been also observed for the coupling constant $C^{\Delta\rho}_{1}$ in Ref.~\cite{hel13}.
From this figure we can argue that the use of very small basis or extra cut-offs  on the $ph$-space in RPA calculations could hide this problem.

It is hard to judge what extrapolation one could make to properly obtain the critical value of the coupling constant at $n \rightarrow \infty$.
For such a reason, we performed an extrapolation by using three functions of the form

\begin{eqnarray}
f_i(n)=\frac{a_i}{n^i}+b_i\,,
\end{eqnarray}

\noindent with $i=1,2,3$. Using the proposed functions, we have set the \emph{lower} limit of the coupling constant by fitting the function $f_1$
and the \emph{upper} limit fitting the function   $f_3$
on the calculated data. The fit done with the function $f_2$
obtains the most reliable critical value, as this function gives the \emph{best} $\chi^2$ value when fitted to the calculated data points for all of the parameterizations of the Skyrme force used in this article. 

\begin{figure}
\begin{center}
\includegraphics[width=0.42\textwidth,angle=-0]{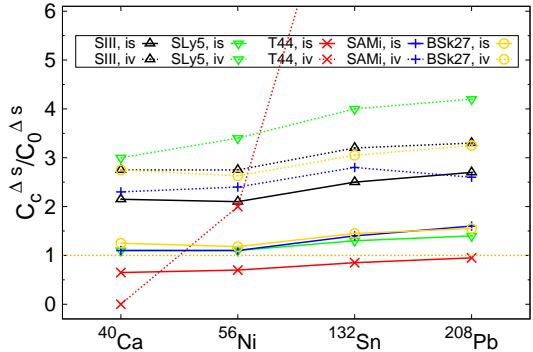}
\end{center}
\caption{(Color online) Ratio among the critical coupling constant and its nominal value using the \emph{best} extrapolation as defined in the text and for the different nuclei considered in our analysis.}
\label{asym}
\end{figure}

Another important aspect is the dependence of the position of the critical coupling constant for the considered nuclei. It is important to notice that as in Ref.~\cite{hel13}, shell effects could hide or enhance the appearance of such a kind of instabilities.
In Fig.~\ref{asym} we represent the ratio $C^{\Delta s}_{0c}/C^{\Delta s}_{0}$ for RPA calculations using the \emph{best} extrapolation and for the four nuclei considered here. 
We notice that in all cases the $N=Z$ nuclei are the most sensitive to the appearance of these instabilities, as discussed in Ref.~\cite{hel13}, the particular shell structure could be at the origin of this phenomenon.

In Table \ref{shell_tab} we summarize the results of our analysis. extrapolation for infinite basis size.

\begin{table}
\caption{Critical parameters, obtained for some Skyrme parameterizations in the limit of infinite number of shells.}
\begin{tabular}{ c|c c c| c c c}
\hline
\hline
 & \multicolumn{3}{c|}{$C^{\Delta s}_{0c}$ (MeV fm)$^5$} & \multicolumn{3}{c}{$C^{\Delta s}_{1c}$ (MeV fm)$^5$} \\[0.5mm]
\hline
& lower & best & higher& lower & best & higher \\[0.5mm]
\hline 
SIII  & 27.28 & 32.37 & 34.05 & 36.55 &43.10 &45.25 \\[0.5mm]
SLy5  & 41.26 & 47.69 & 49.80 & 36.90 &43.47 &45.63 \\[0.5mm]
T44   & 39.06 & 47.28 & 49.96 &-13.02 &-2.01 & 1.60 \\[0.5mm]
BSk27 & 22.11 & 28.22 & 30.05 & 33.16 &38.81 &40.67 \\[0.5mm]
SAMi  & 41.30 & 47.54 & 49.58 & 38.81 &45.15 &47.23 \\ [0.5mm]
\hline
\hline
\end{tabular}
\label{shell_tab}
\end{table}

\begin{figure}
\begin{center}
\includegraphics[width=0.42\textwidth,angle=-0]{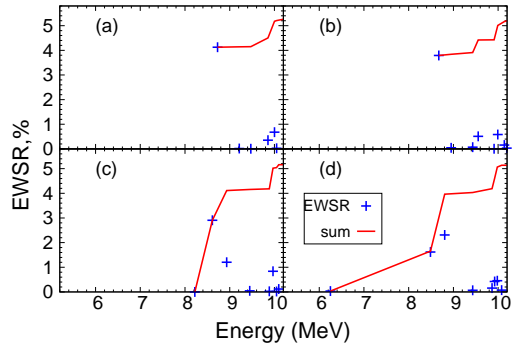}
\end{center}
\caption{(Color online) Evolution of the low-lying spectrum in $^{132}$Sn (calculated with 16 oscillator shells) with respect to $C^{\Delta s}_{0}$ for the SLy5 functional. In panel (a) we present the spectrum calculated using $C^{\Delta s}_{0}$=9.22 MeV fm$^5$, panel (b) $C^{\Delta s}_{0}$=46.10 MeV fm$^5$, panel (c)  $C^{\Delta s}_{0}$=55.3 MeV fm$^5$ and panel (d)  $C^{\Delta s}_{0}$= 59.91 MeV fm$^5$.}
\label{pygmy}
\end{figure}

In our analysis, we pay particular attention to the behavior of the low lying dipole excitations, since recent results~\cite{suz90} show the appearance of pigmy dipole resonances in atomic nuclei. Moreover low-lying dipole states play an important role in describing some relevant astrophysical process.  We recall, that in our calculations, all the RPA phonons in the $J=1^{-}$ channel have been orthonormalized against the spurious state by construction as explained in Refs.~\cite{nak07,toi10}, thus our results are not polluted by spuriosity.
Varying the isoscalar coupling constant, we have calculated the strength concentrated in the low-lying states in $^{132}$Sn using the SLy5 parameterization  and different nominal values for the isoscalar coupling constant $C^{\Delta s}_{0}$. In Fig.\ref{pygmy} we show the results. It is clearly seen that when the value of the coupling constant gets  close to its critical value,  we observe a noticeable shift of the strength  to lower energies, and there is some fragmentation of this strength over several states. This result seems to be artificial and marks a possible weakness for some calculations performed for the distribution of the low-lying strength using functionals with coupling constants that are close to their critical value.
It is worth emphasizing that once we are above the critical value, the energy weighted sum rule can not be fulfilled anymore since some of the strength is taken away by the imaginary phonons.

Finally we stress that even though some numerical effects could influence the precision of our calculations, the effect of collapsing solutions in a very tight region of values, could be reproduced using different approaches realized in different numerical  codes.
In particular the results obtained with the present RPA code has been checked~\cite{kor15} using another RPA code~\cite{sto11} based on Finite Amplitude Method~\cite{nak07}.

\section{RPA instabilities in infinite matter}\label{rpa:snm}

In this section, we present the formalism of the linear response theory in infinite nuclear matter. This method has been the subject of a recent  review article~\cite{pas14D}, where all the details of the formalism are presented and discussed. We limit ourselves to sketch the basic ingredients of the formalism.

The first ingredient is the Hartree-Fock retarded propagator of a non-interacting $ph$-pair
Since for the present study we ignore charge-exchange process, the particle and the hole in the same pair share the same isospin number $\tau=p,n$

\begin{eqnarray}
G^{(\tau)}_{HF}(\mathbf{k},\mathbf{p},\omega)=\frac{\theta(k_{F}^{\tau}-k)-\theta(k_{F}^{\tau}-|\mathbf{k}+\mathbf{q}|)}{\omega+\varepsilon_{\tau}(\mathbf{k})-\varepsilon_{\tau}(\mathbf{k}+\mathbf{q})+i\eta}\,,
\end{eqnarray}

\noindent where $\theta(k_{F}^{\tau}-k)$ is the standard step-function and $\varepsilon_{\tau}(\mathbf{k})=\frac{\hbar^{2}\mathbf{k}^{2}}{2m^{*}_{\tau}}+U_{\tau}$ is the single particle energy and $m^{*}_{\tau},U_{\tau}$ represent the effective mass and the single particle potential, respectively, while $\mathbf{k}$ is the moment of the hole and $\mathbf{q}$ the external momentum transferred by the probe we use to excite the system. The latter can be taken along the $z$-axis without loss of generality~\cite{gar92}.
For simplicity, we will illustrate the case of SNM, but the formalism has been already generalized to the more general case of isospin asymmetric nuclear matter (ANM)~\cite{pas14D,dav14A}.
Since the two Fermi surfaces are equal in SNM, we can drop the $\tau$ index. The correlated RPA propagator is obtained by solving the Bethe-Salpeter (BS) equations as

\begin{figure*}[!h]
\begin{center}
\includegraphics[width=0.42\textwidth,angle=-90]{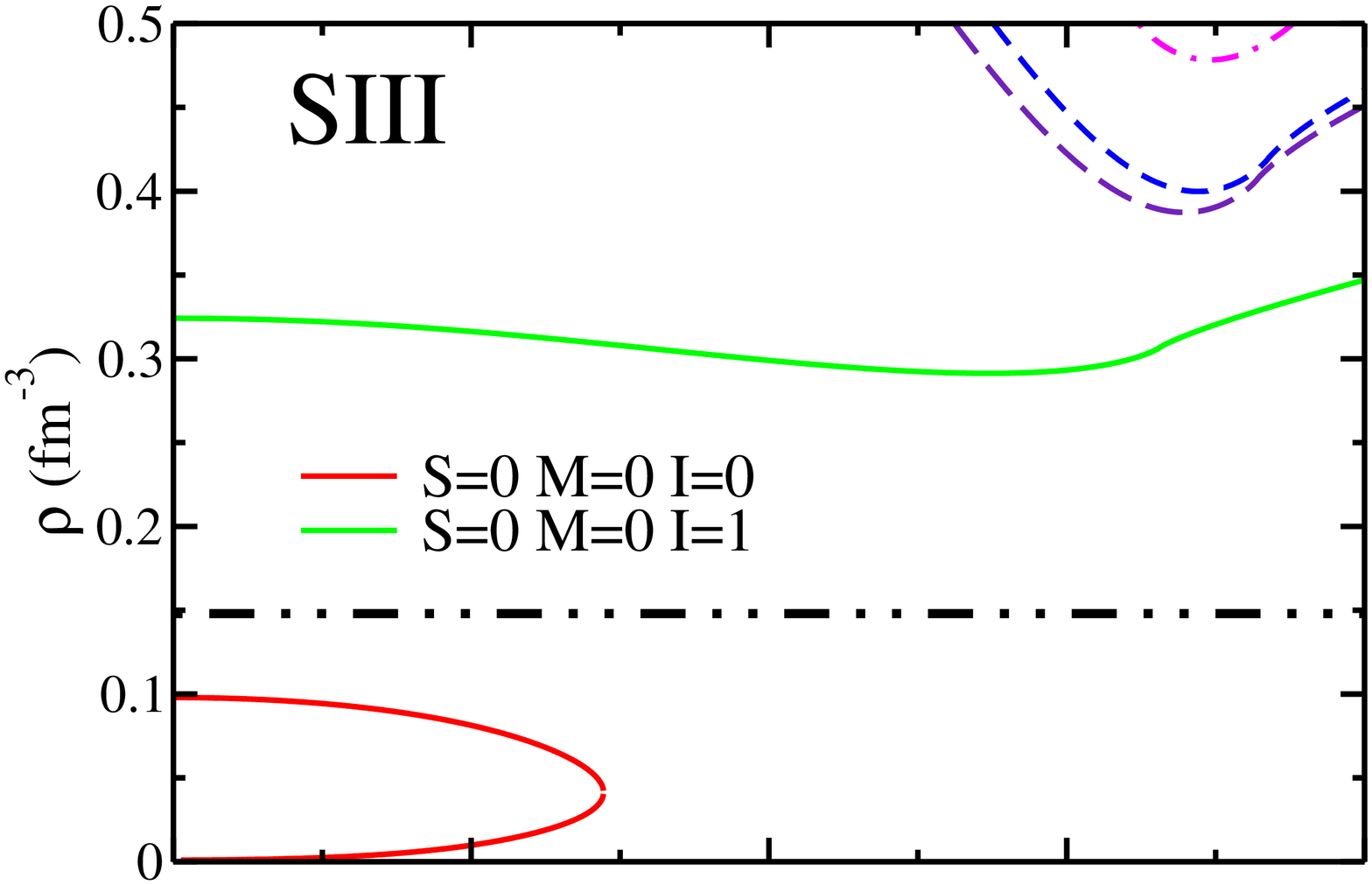}
\hspace{-2.47cm}
\includegraphics[width=0.42\textwidth,angle=-90]{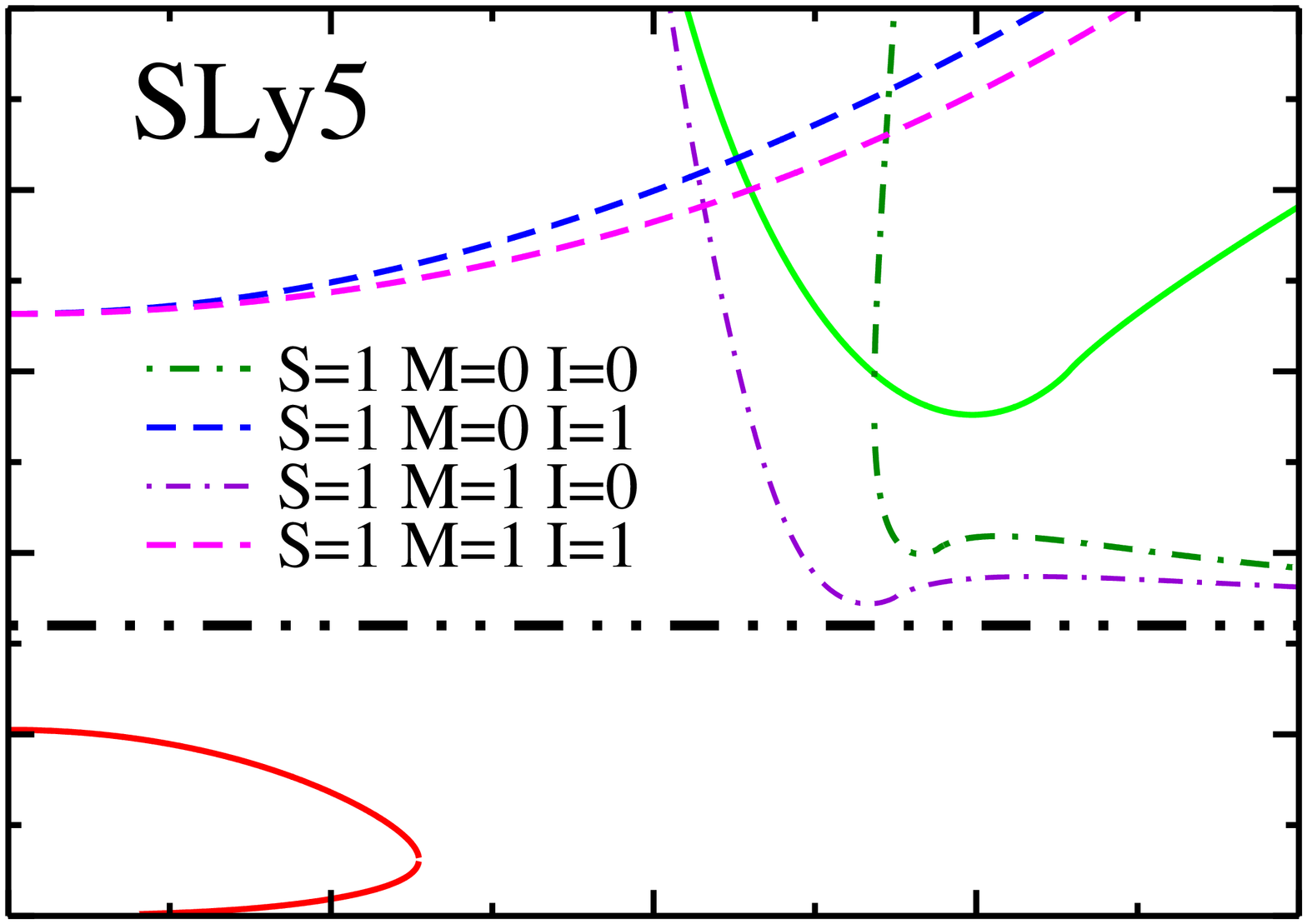}\\
\vspace{-2.3cm}
\includegraphics[width=0.42\textwidth,angle=-90]{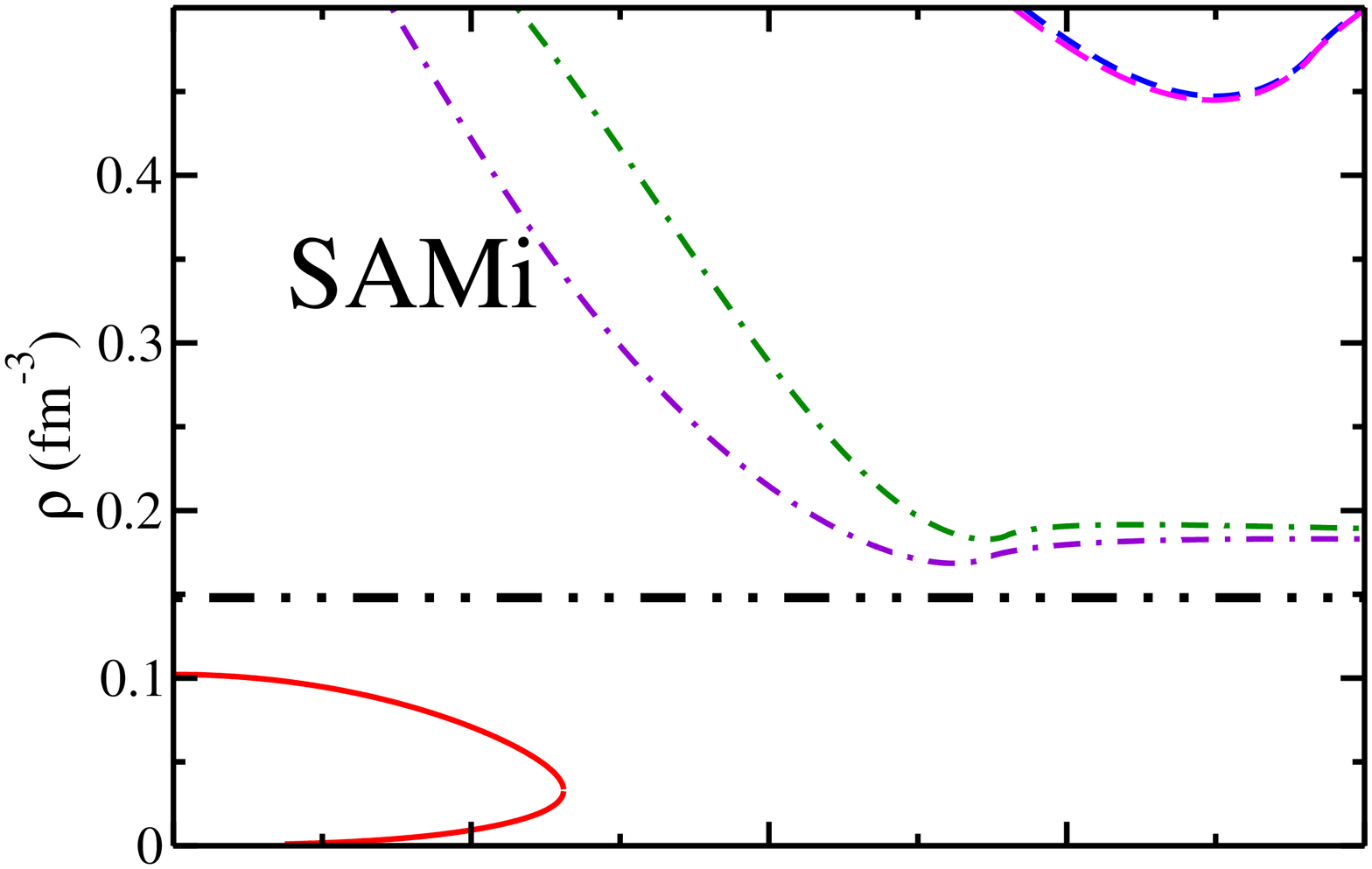}
\hspace{-2.47cm}
\includegraphics[width=0.42\textwidth,angle=-90]{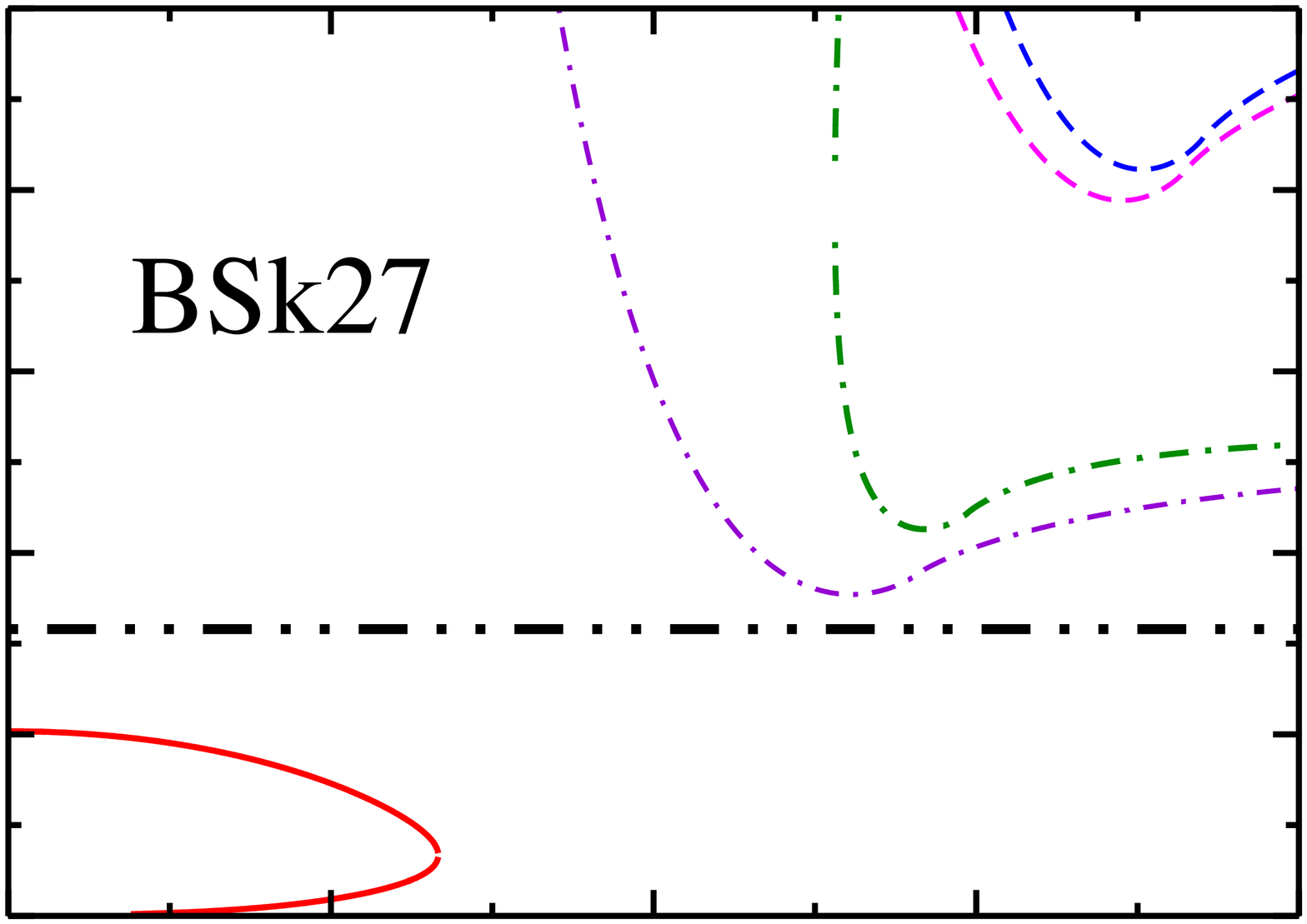}\\
\vspace{-2.3cm}
\includegraphics[width=0.42\textwidth,angle=-90]{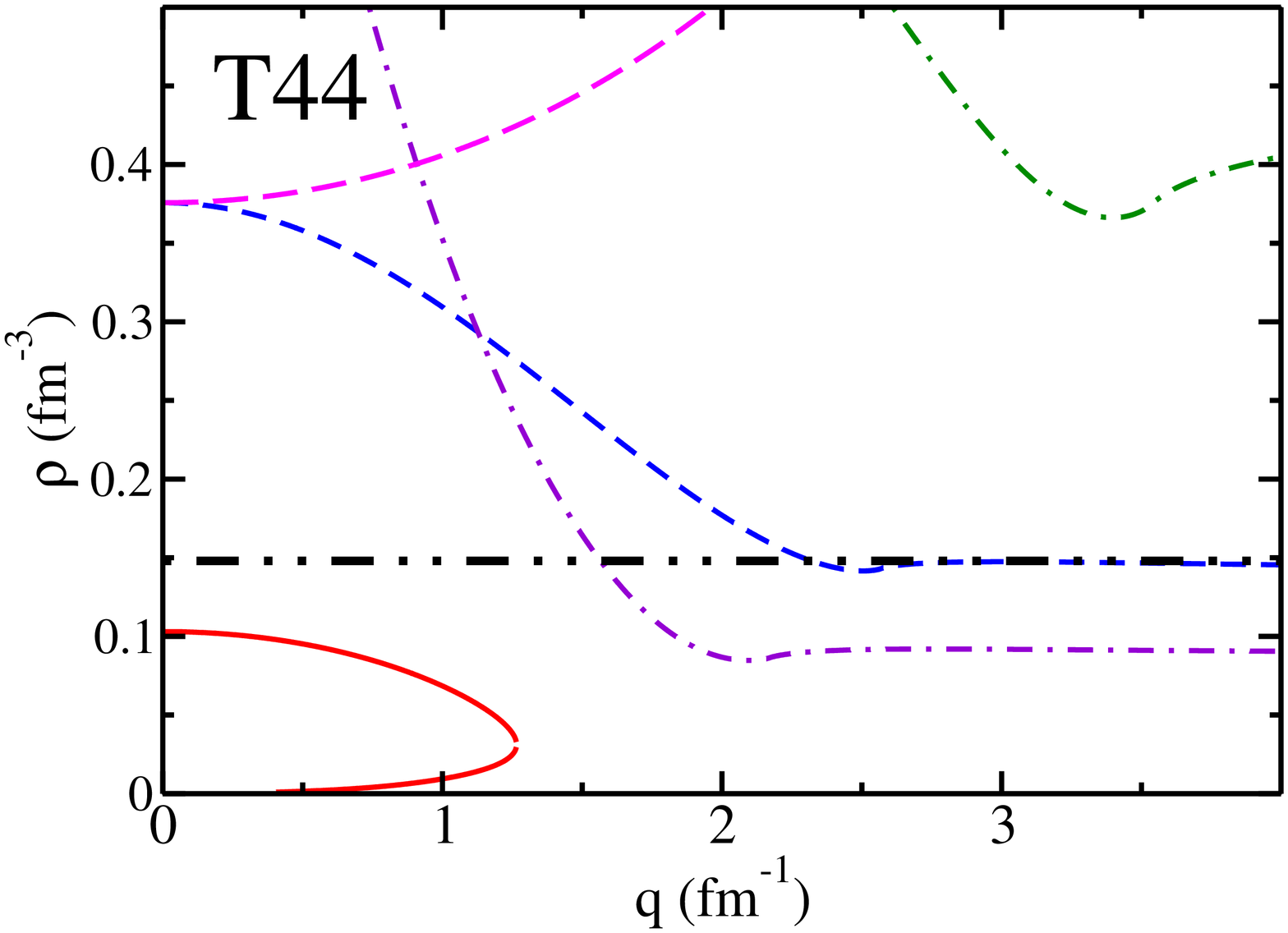}
\hspace{-2.47cm}
\includegraphics[width=0.42\textwidth,angle=-90]{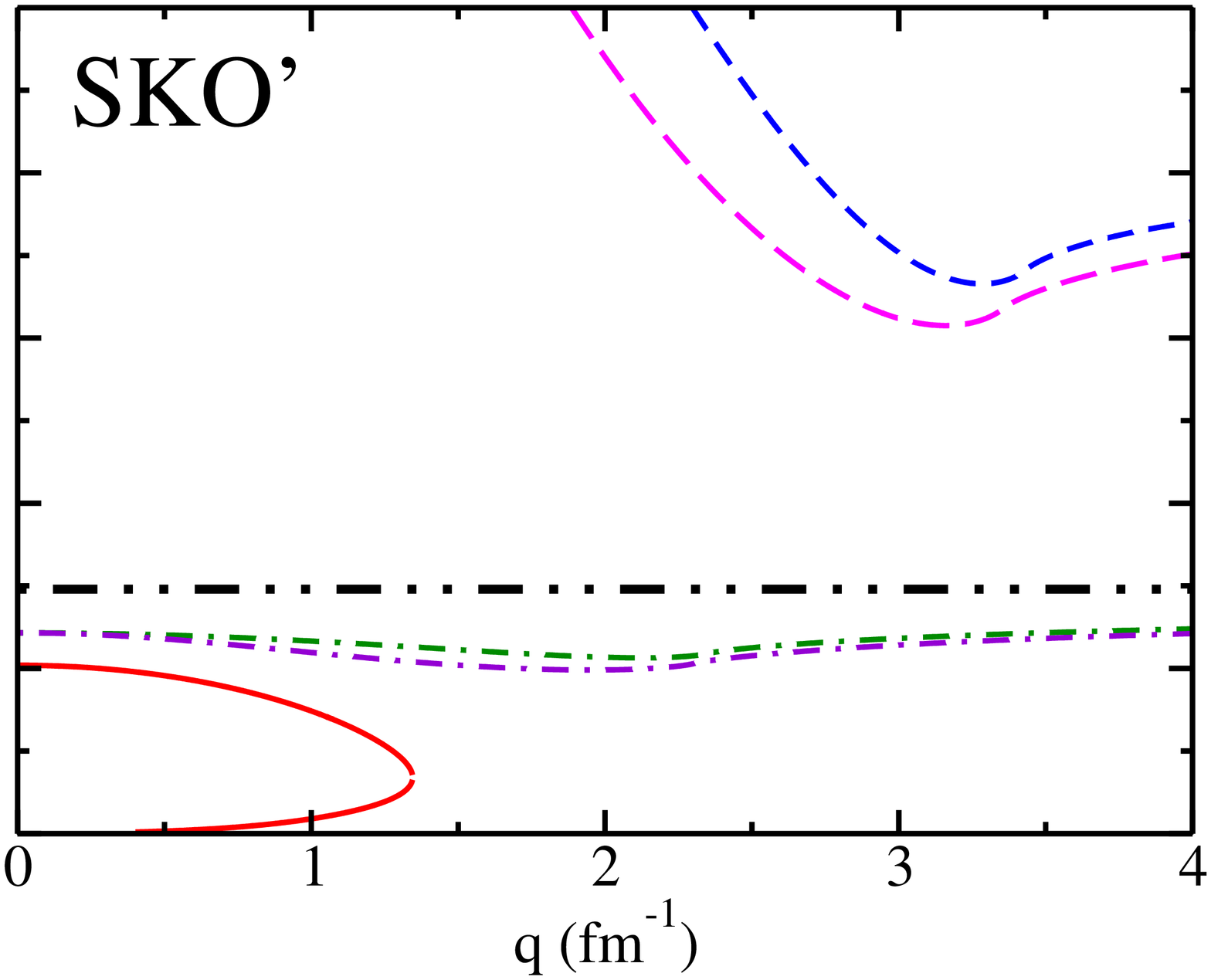}
\end{center}
\caption{(Color online) Instabilities in SNM for the functionals considered in the present article. The dashed-dotted horizontal line stands for the saturation density of the functional. See text for details.}
\label{poles}
\end{figure*}

\begin{widetext}
\begin{eqnarray}
G^{(\alpha)}(\mathbf{k}_{1},q,\omega)=\delta_{\alpha\alpha'}G_{HF}(\mathbf{k}_{1},q,\omega)+G_{HF}(\mathbf{k}_{1},q,\omega)\sum_{\alpha'}\int \frac{d^{3}\mathbf{k}_{2}}{2\pi^{3}}V_{ph}^{(\alpha,\alpha')}(\mathbf{k}_{1},\mathbf{k}_{2})G^{\alpha'}(\mathbf{k}_{2},q,\omega)\,,
\end{eqnarray}
\end{widetext}

\noindent where $\alpha\equiv(S,M,I)$ is a shorthand notation which stands for the quantum number of the system: the total spin $S$ and its projection $M$ along the $z$-axis and the total isospin $I$.
Thanks to the particular form of the residual interaction  $V_{ph}^{(\alpha,\alpha')}$, we can solve the BS equation analytically. The linear response function can be then obtained by integrating over the hole momentum $\mathbf{k}_{1}$ as

\begin{eqnarray}
\chi^{(\alpha)}(q,\omega)=n_{d}\langle G^{(\alpha)}(\mathbf{k}_{1},q,\omega) \rangle\,,
\end{eqnarray}

\noindent where $n_{d}=4$ is the degeneracy of the system. The explicit expressions of $\chi^{\alpha}(q,\omega)$ are quite complicated and we refer the reader to Ref.~\cite{pas14D} for their explicit expressions.
To detect a pole we can simply look for the zeros of the expression
\begin{eqnarray}\label{eq:poles}
1/\chi^{(\alpha)}(\omega=0,q)=0\,,
\end{eqnarray}
\noindent for given values of transferred momentum $q$ and density $\rho$.
In Fig.~\ref{poles}, we show in the $(\rho,q)$-plane the position of the instabilities  obtained solving Eq.~(\ref{eq:poles}).
Apart from the spinodal instability which is related to the gas-liquid phase transition~\cite{duc08} and manifests itself in the low-density region in the (0,0,0) channel, all the other instabilities are unphysical.
In particular, we observe that all the functionals apart from SIII, present an instability in the spin channel close to the saturation density.
Such an instability indicates a phase transition between spin-unpolarized to spin-polarized symmetric matter.
In the present article, we investigate how these ferromagnetic instabilities are related to the problems previously detected in finite nuclei.

\begin{figure}
\begin{center}
\includegraphics[width=0.42\textwidth,angle=-90]{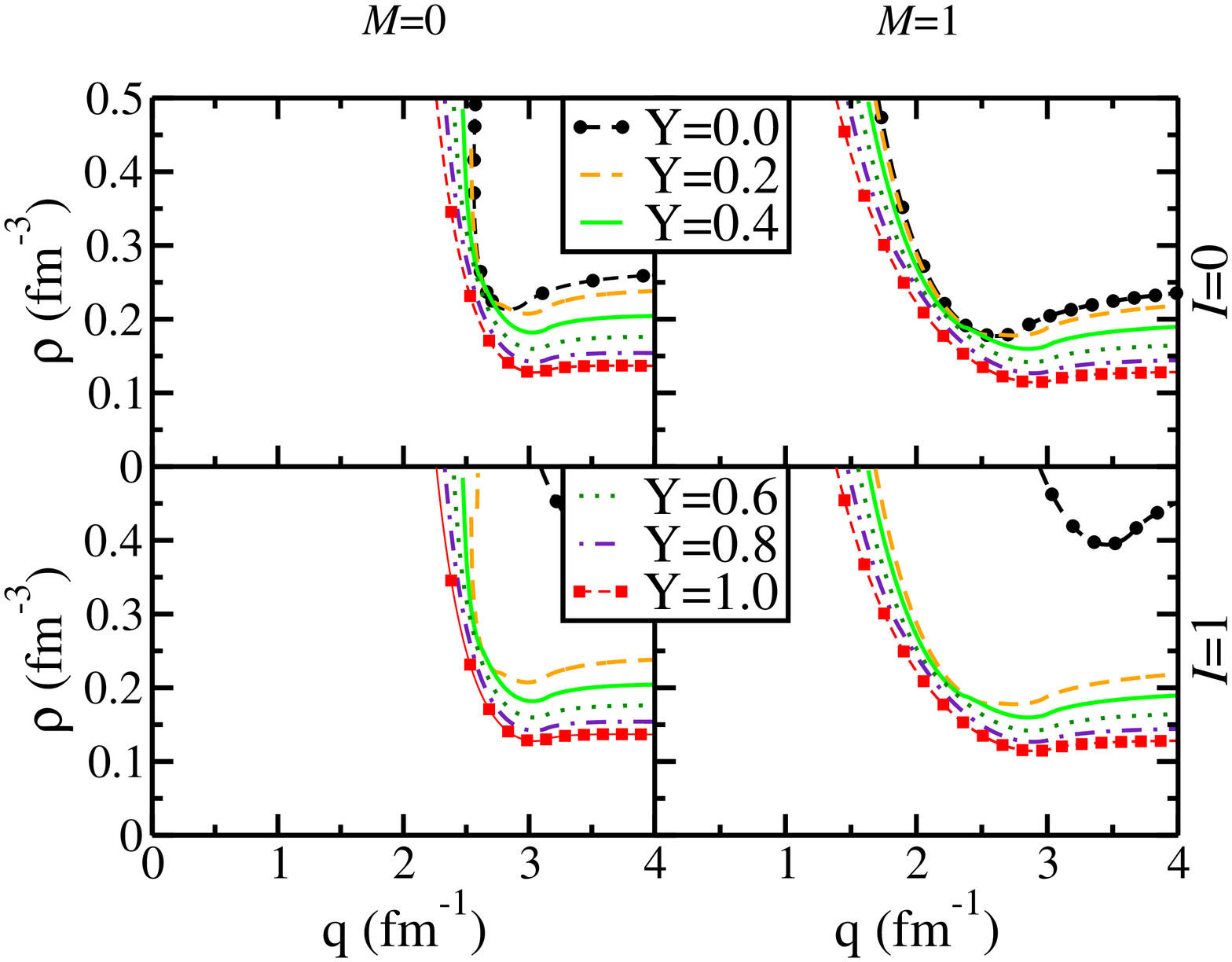}
\end{center}
\caption{(Color online) Evolution of the critical densities in the S=1 channel in asymmetric matter for different values of the asymmetry parameter $Y$ for the BSk27 functional. }
\label{pole:asym}
\end{figure}

As already discussed in Ref.~\cite{dav14A}, the position of the poles can be different when we vary the isospin asymmetry of the system.
In Fig.~\ref{pole:asym}, we show the evolution of the poles in asymmetric matter in the $S=1$ channel and different asymmetries for the BSk27 functional.  
We observe that changing the asymmetry parameter $Y=\frac{\rho_n-\rho_p}{\rho_n+\rho_p}$  from 0 ($i.e.$ SNM) to 0.2 (the bulk asymmetry of $^{208}$Pb) the vector-isovector instability gets closer to saturation density, and one could thus expect the appearance of instabilities in very neutron rich nuclei.
As one can see from Fig.~\ref{pole:asym}, the variation of the poles is roughly monotonic from SNM to PNM ($i.e.$  $Y=1$) (see also discussion in Ref.~\cite{pas14D}). We remind that in this case both the isoscalar and isovector coupling constants keep their nominal value. For such a reason, this result can not directly compared with the one obtained in Fig.\ref{asym}. Anticipating the result of the next section, we conclude that the stability criterion derived here should be applied for both SNM and PNM as discussed in Ref.~\cite{pas13}.

\section{Detecting instabilities}\label{sec:nuclei}

\begin{figure}
\begin{center}
\includegraphics[width=0.42\textwidth,angle=-90]{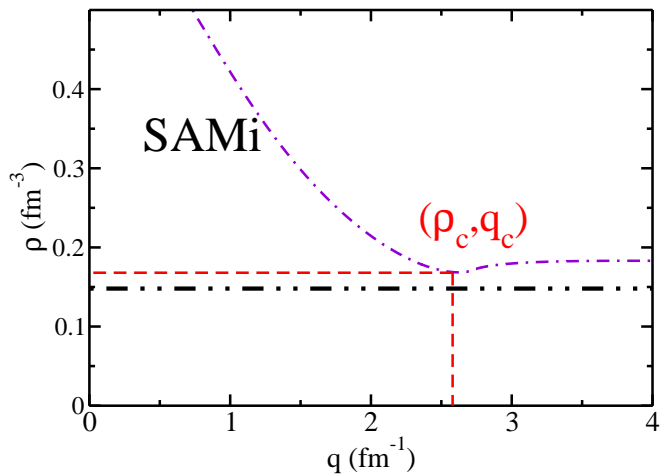}
\end{center}
\caption{(Color online) Position of the poles in the $(\rho,q)$-plane. The minimum of this curve defines the point $(\rho_{c},q_{c})$.}
\label{Extr}
\end{figure}

In Fig.~\ref{Extr}, we illustrate the process used to extract the critical density $\rho_{c}$ at which a pole occurs in SNM: we consider the lowest density value at which the pole occurs in a given spin-isospin channel. Although  we are not presently interested in knowing the associated value of the transferred momentum $q_{c}$ for a given $\rho_c$, it is anyhow useful to distinguish between long-wavelength and finite-size instabilities. Since we consider functionals originated from zero-range Skyrme interactions, there is no upper limit to the value of the exchange momentum between $ph$ pairs.

In Tab.~\ref{tab:sky:crit}, we report the values of the critical densities using the coupling constants extracted from Tab.~\ref{shell_tab}.
Each value of $\rho_{c}$ is obtained using the \emph{best} value of the critical coupling constant extracted in Tab.\ref{shell_tab}, the error bars are extracted using the \emph{higher} and \emph{lower} value respectively for both isoscalar and isovector channels.

\begin{table}
\begin{center}
\caption{Critical coupling constant and critical density for the functionals studied in the present article extracted at $n \rightarrow \infty $.  The error bars we have added come from the uncertainty in the extrapolation we used for studying the variation of the critical point as a function of the shell number. For each interaction we also give the position of the pole for the nominal value of the coupling constant.}
\begin{tabular}{c|cc|cc}
\hline
\hline
         & \multicolumn{2}{c|}{isoscalar}                      &  \multicolumn{2}{c}{isovector    }         \\
         &   $\rho_{c}/\rho_{sat}$&  $\rho_{pole}/\rho_{sat}$&$\rho_{c}/\rho_{sat}$& $\rho_{pole}/\rho_{sat}$\\[1mm]
\hline
  SIII & $1.13^{+0.34}_{-0.06}$ &3.29 &$1.05^{+0.20}_{-0.06}$ &2.66\\[1mm]
  SLy5& $1.01^{+0.26}_{-0.06}$&1.07 &$0.97^{+0.20}_{-0.05}$ &2.07\\ [1mm]
    T44 & $0.96^{+0.26}_{-0.07}$ & 0.52&$0.96^{+0.37}_{-0.08}$&0.88\\[1mm]
  BSk27& $1.11^{+0.33}_{-0.10}$  & 1.12&$1.00_{-0.04}^{+0.19}$&2.48\\[1mm]
  SAMi & $0.97_{-0.06}^{+0.21}$ & 1.06&$1.00_{-0.05}^{+0.24}$ &2.80\\[1mm]
  \hline 
  \hline
\end{tabular}
\label{tab:sky:crit}
\end{center}
\end{table}

\begin{figure}
\begin{center}
\includegraphics[width=0.42\textwidth,angle=-90]{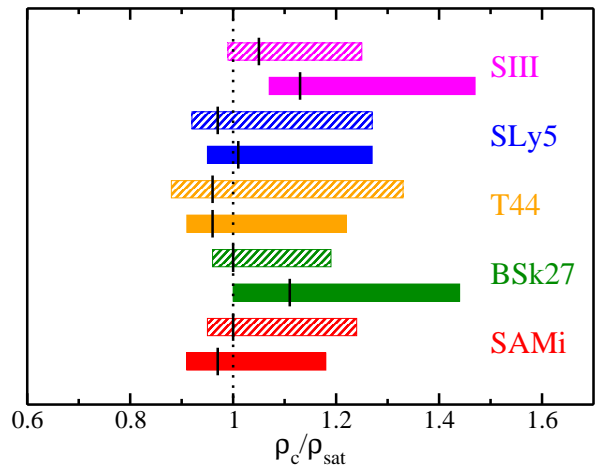}
\end{center}
\caption{(Color online) Critical density $\rho_{c}/\rho_{sat}$. The uncertainty  band comes from the numerical extraction and the number of shells employed in the calculations. Solid bands refer to the isoscalar channel, while dashed bands refer to the isovector one. The vertical black lines refer to the \emph{best} value as given in Tab.\ref{tab:sky:crit}}
\label{bands}
\end{figure}

This value is shown in  Fig.\ref{bands} by means of a vertical solid black line in both isoscalar and isovector channels.
We observe that most of the error bars fall in a range of the density $\rho_c\in[\rho_{sat},1.3\rho_{sat}]$. Compared to Ref.~\cite{hel13}, it seems that performing RPA calculations allow us to explore slightly higher density regions.

We can conclude that the stability criterion derived in Ref.~\cite{hel13} is also valid for the vector part of the functional and thus can be used to detect systematically this kind of problems.
We also stress that the present method can be easily incorporated into a fitting procedure to avoid by construction this kind of problems, but in this case it is better to consider safer criteria by excluding a larger density interval.

\section{Conclusions}\label{sec:conclusion}

We have investigated the appearance of finite-size instabilities in RPA calculations of finite nuclei. The instabilities manifest trough the appearance of imaginary phonons and by an anomalous accumulation of strength in the low-energy part of the vibrational spectrum.
By using six representative Skyrme functionals, we have identified the terms of the functional which are a possible source of these instabilities and given upper limits to the values of their coupling constants.
We have also performed systematic RPA calculations in symmetric nuclear matter and we have observed that the presence of such a kind of instabilities can be related to the appearance of spurious phase transitions in the infinite medium around saturation density. As done in Ref.~\cite{hel13}, we have thus extracted a quantitative criterion to detect the  presence of finite-size instabilities. Since LR calculations in SNM and PNM are not time consuming, such a method can be easily included into a fitting protocol, thus avoiding such kind of problems by construction. An example  has been already presented in Ref.~\cite{pas13}. 

Although the  finite range interactions do not seem to manifest such kind of pathologies since the finite range works as an effective cut-off in momentum space. It would be probably interesting in the future to do a similar analysis especially for the functionals including explicit tensor terms~\cite{nak03,ang12,gra13}, which have been source of several instabilities in the Skyrme case~\cite{pas11}.

\section*{Acknowledgments}

The authors thank T. Duguet for the stimulating discussions which motivated the present article. We also acknowledge the fruitful discussions with K. Bennaceur, J. Dobaczewski and M. Kortelainen. A.P. is also grateful to the University of Jyv\"askyl\"a and the FIDIPRO group for the warm hospitality during which part of this work has been realized.
J.N. has been supported by Mineco (Spain), grant FIS2014-51948-C2-1-P. D.T. has been supported in part by the ERANET-
NuPNET grant SARFEN of the Polish National Centre for Research and Development (NCBiR) and by the Academy of Finland and University of Jyv\"askyl\"a within the FIDIPRO program


\bibliography{biblio}

\begin{thebibliography}{50}
\expandafter\ifx\csname natexlab\endcsname\relax\def\natexlab#1{#1}\fi
\expandafter\ifx\csname bibnamefont\endcsname\relax
  \def\bibnamefont#1{#1}\fi
\expandafter\ifx\csname bibfnamefont\endcsname\relax
  \def\bibfnamefont#1{#1}\fi
\expandafter\ifx\csname citenamefont\endcsname\relax
  \def\citenamefont#1{#1}\fi
\expandafter\ifx\csname url\endcsname\relax
  \def\url#1{\texttt{#1}}\fi
\expandafter\ifx\csname urlprefix\endcsname\relax\def\urlprefix{URL }\fi
\providecommand{\bibinfo}[2]{#2}
\providecommand{\eprint}[2][]{\url{#2}}

\bibitem[{\citenamefont{Bender et~al.}(2003)\citenamefont{Bender, Heenen, and
  Reinhard}}]{ben03}
\bibinfo{author}{\bibfnamefont{M.}~\bibnamefont{Bender}},
  \bibinfo{author}{\bibfnamefont{P.-H.} \bibnamefont{Heenen}},
  \bibnamefont{and} \bibinfo{author}{\bibfnamefont{P.-G.}
  \bibnamefont{Reinhard}}, \bibinfo{journal}{Rev. Mod. Phys.}
  \textbf{\bibinfo{volume}{75}}, \bibinfo{pages}{121} (\bibinfo{year}{2003}).

\bibitem[{\citenamefont{Erler et~al.}(2012)\citenamefont{Erler, Birge,
  Kortelainen, Nazarewicz, Olsen, Perhac, and Stoitsov}}]{erl12}
\bibinfo{author}{\bibfnamefont{J.}~\bibnamefont{Erler}},
  \bibinfo{author}{\bibfnamefont{N.}~\bibnamefont{Birge}},
  \bibinfo{author}{\bibfnamefont{M.}~\bibnamefont{Kortelainen}},
  \bibinfo{author}{\bibfnamefont{W.}~\bibnamefont{Nazarewicz}},
  \bibinfo{author}{\bibfnamefont{E.}~\bibnamefont{Olsen}},
  \bibinfo{author}{\bibfnamefont{A.~M.} \bibnamefont{Perhac}},
  \bibnamefont{and} \bibinfo{author}{\bibfnamefont{M.}~\bibnamefont{Stoitsov}},
  \bibinfo{journal}{Nature} \textbf{\bibinfo{volume}{486}},
  \bibinfo{pages}{509} (\bibinfo{year}{2012}).

\bibitem[{\citenamefont{Kortelainen et~al.}(2010)\citenamefont{Kortelainen,
  Lesinski, Mor\'e, Nazarewicz, Sarich, Schunck, Stoitsov, and Wild}}]{kor10}
\bibinfo{author}{\bibfnamefont{M.}~\bibnamefont{Kortelainen}},
  \bibinfo{author}{\bibfnamefont{T.}~\bibnamefont{Lesinski}},
  \bibinfo{author}{\bibfnamefont{J.}~\bibnamefont{Mor\'e}},
  \bibinfo{author}{\bibfnamefont{W.}~\bibnamefont{Nazarewicz}},
  \bibinfo{author}{\bibfnamefont{J.}~\bibnamefont{Sarich}},
  \bibinfo{author}{\bibfnamefont{N.}~\bibnamefont{Schunck}},
  \bibinfo{author}{\bibfnamefont{M.~V.} \bibnamefont{Stoitsov}},
  \bibnamefont{and} \bibinfo{author}{\bibfnamefont{S.}~\bibnamefont{Wild}},
  \bibinfo{journal}{Phys. Rev. C} \textbf{\bibinfo{volume}{82}},
  \bibinfo{pages}{024313} (\bibinfo{year}{2010}).

\bibitem[{\citenamefont{Kortelainen et~al.}(2012)\citenamefont{Kortelainen,
  McDonnell, Nazarewicz, Reinhard, Sarich, Schunck, Stoitsov, and
  Wild}}]{kor12}
\bibinfo{author}{\bibfnamefont{M.}~\bibnamefont{Kortelainen}},
  \bibinfo{author}{\bibfnamefont{J.}~\bibnamefont{McDonnell}},
  \bibinfo{author}{\bibfnamefont{W.}~\bibnamefont{Nazarewicz}},
  \bibinfo{author}{\bibfnamefont{P.-G.} \bibnamefont{Reinhard}},
  \bibinfo{author}{\bibfnamefont{J.}~\bibnamefont{Sarich}},
  \bibinfo{author}{\bibfnamefont{N.}~\bibnamefont{Schunck}},
  \bibinfo{author}{\bibfnamefont{M.~V.} \bibnamefont{Stoitsov}},
  \bibnamefont{and} \bibinfo{author}{\bibfnamefont{S.~M.} \bibnamefont{Wild}},
  \bibinfo{journal}{Phys. Rev. C} \textbf{\bibinfo{volume}{85}},
  \bibinfo{pages}{024304} (\bibinfo{year}{2012}).

\bibitem[{\citenamefont{Kortelainen et~al.}(2014)\citenamefont{Kortelainen,
  McDonnell, Nazarewicz, Olsen, Reinhard, Sarich, Schunck, Wild, Davesne, Erler
  et~al.}}]{kor13}
\bibinfo{author}{\bibfnamefont{M.}~\bibnamefont{Kortelainen}},
  \bibinfo{author}{\bibfnamefont{J.}~\bibnamefont{McDonnell}},
  \bibinfo{author}{\bibfnamefont{W.}~\bibnamefont{Nazarewicz}},
  \bibinfo{author}{\bibfnamefont{E.}~\bibnamefont{Olsen}},
  \bibinfo{author}{\bibfnamefont{P.-G.} \bibnamefont{Reinhard}},
  \bibinfo{author}{\bibfnamefont{J.}~\bibnamefont{Sarich}},
  \bibinfo{author}{\bibfnamefont{N.}~\bibnamefont{Schunck}},
  \bibinfo{author}{\bibfnamefont{S.~M.} \bibnamefont{Wild}},
  \bibinfo{author}{\bibfnamefont{D.}~\bibnamefont{Davesne}},
  \bibinfo{author}{\bibfnamefont{J.}~\bibnamefont{Erler}},
  \bibnamefont{et~al.}, \bibinfo{journal}{Phys. Rev. C}
  \textbf{\bibinfo{volume}{89}}, \bibinfo{pages}{054314}
  (\bibinfo{year}{2014}).

\bibitem[{\citenamefont{Chabanat et~al.}(1997)\citenamefont{Chabanat, Bonche,
  Haensel, Meyer, and Schaeffer}}]{chab97}
\bibinfo{author}{\bibfnamefont{E.}~\bibnamefont{Chabanat}},
  \bibinfo{author}{\bibfnamefont{P.}~\bibnamefont{Bonche}},
  \bibinfo{author}{\bibfnamefont{P.}~\bibnamefont{Haensel}},
  \bibinfo{author}{\bibfnamefont{J.}~\bibnamefont{Meyer}}, \bibnamefont{and}
  \bibinfo{author}{\bibfnamefont{R.}~\bibnamefont{Schaeffer}},
  \bibinfo{journal}{Nucl. Phys. A} \textbf{\bibinfo{volume}{627}},
  \bibinfo{pages}{710} (\bibinfo{year}{1997}).

\bibitem[{\citenamefont{Washiyama et~al.}(2012)\citenamefont{Washiyama,
  Bennaceur, Avez, Bender, Heenen, and Hellemans}}]{was12}
\bibinfo{author}{\bibfnamefont{K.}~\bibnamefont{Washiyama}},
  \bibinfo{author}{\bibfnamefont{K.}~\bibnamefont{Bennaceur}},
  \bibinfo{author}{\bibfnamefont{B.}~\bibnamefont{Avez}},
  \bibinfo{author}{\bibfnamefont{M.}~\bibnamefont{Bender}},
  \bibinfo{author}{\bibfnamefont{P.-H.} \bibnamefont{Heenen}},
  \bibnamefont{and}
  \bibinfo{author}{\bibfnamefont{V.}~\bibnamefont{Hellemans}},
  \bibinfo{journal}{Phys. Rev. C} \textbf{\bibinfo{volume}{86}},
  \bibinfo{pages}{054309} (\bibinfo{year}{2012}).

\bibitem[{\citenamefont{Chappert et~al.}(2015)\citenamefont{Chappert, Pillet,
  Girod, and Berger}}]{cha15}
\bibinfo{author}{\bibfnamefont{F.}~\bibnamefont{Chappert}},
  \bibinfo{author}{\bibfnamefont{N.}~\bibnamefont{Pillet}},
  \bibinfo{author}{\bibfnamefont{M.}~\bibnamefont{Girod}}, \bibnamefont{and}
  \bibinfo{author}{\bibfnamefont{J.-F.} \bibnamefont{Berger}},
  \bibinfo{journal}{Phys. Rev. C} \textbf{\bibinfo{volume}{91}},
  \bibinfo{pages}{034312} (\bibinfo{year}{2015}).

\bibitem[{\citenamefont{Dobaczewski et~al.}(2014)\citenamefont{Dobaczewski,
  Nazarewicz, and Reinhard}}]{dob14}
\bibinfo{author}{\bibfnamefont{J.}~\bibnamefont{Dobaczewski}},
  \bibinfo{author}{\bibfnamefont{W.}~\bibnamefont{Nazarewicz}},
  \bibnamefont{and} \bibinfo{author}{\bibfnamefont{P.}~\bibnamefont{Reinhard}},
  \bibinfo{journal}{J. Phys. G: Nucl. Part. Phys.}
  \textbf{\bibinfo{volume}{41}}, \bibinfo{pages}{074001}
  (\bibinfo{year}{2014}).

\bibitem[{\citenamefont{Erler et~al.}(2010)\citenamefont{Erler, Kl{\"u}pfel,
  and Reinhard}}]{erl10}
\bibinfo{author}{\bibfnamefont{J.}~\bibnamefont{Erler}},
  \bibinfo{author}{\bibfnamefont{P.}~\bibnamefont{Kl{\"u}pfel}},
  \bibnamefont{and} \bibinfo{author}{\bibfnamefont{P.}~\bibnamefont{Reinhard}},
  \bibinfo{journal}{J. Phys. G: Nucl. Part. Phys.}
  \textbf{\bibinfo{volume}{37}}, \bibinfo{pages}{064001}
  (\bibinfo{year}{2010}).

\bibitem[{\citenamefont{Margueron et~al.}(2002)\citenamefont{Margueron,
  Navarro, and Van~Giai}}]{mar02}
\bibinfo{author}{\bibfnamefont{J.}~\bibnamefont{Margueron}},
  \bibinfo{author}{\bibfnamefont{J.}~\bibnamefont{Navarro}}, \bibnamefont{and}
  \bibinfo{author}{\bibfnamefont{N.}~\bibnamefont{Van~Giai}},
  \bibinfo{journal}{Phys. Rev. C} \textbf{\bibinfo{volume}{66}},
  \bibinfo{pages}{014303} (\bibinfo{year}{2002}).

\bibitem[{\citenamefont{Navarro and Polls}(2013)}]{nav13}
\bibinfo{author}{\bibfnamefont{J.}~\bibnamefont{Navarro}} \bibnamefont{and}
  \bibinfo{author}{\bibfnamefont{A.}~\bibnamefont{Polls}},
  \bibinfo{journal}{Phys. Rev. C} \textbf{\bibinfo{volume}{87}},
  \bibinfo{pages}{044329} (\bibinfo{year}{2013}).

\bibitem[{\citenamefont{Davesne
  et~al.}(2014{\natexlab{a}})\citenamefont{Davesne, Pastore, and
  Navarro}}]{dav14F}
\bibinfo{author}{\bibfnamefont{D.}~\bibnamefont{Davesne}},
  \bibinfo{author}{\bibfnamefont{A.}~\bibnamefont{Pastore}}, \bibnamefont{and}
  \bibinfo{author}{\bibfnamefont{J.}~\bibnamefont{Navarro}},
  \bibinfo{journal}{J. Phys. G: Nucl.Part. Phys.}
  \textbf{\bibinfo{volume}{41}}, \bibinfo{pages}{065104}
  (\bibinfo{year}{2014}{\natexlab{a}}).

\bibitem[{\citenamefont{Chamel and Goriely}(2010)}]{cha10}
\bibinfo{author}{\bibfnamefont{N.}~\bibnamefont{Chamel}} \bibnamefont{and}
  \bibinfo{author}{\bibfnamefont{S.}~\bibnamefont{Goriely}},
  \bibinfo{journal}{Phys. Rev. C} \textbf{\bibinfo{volume}{82}},
  \bibinfo{pages}{045804} (\bibinfo{year}{2010}).

\bibitem[{\citenamefont{Lesinski et~al.}(2006)\citenamefont{Lesinski,
  Bennaceur, Duguet, and Meyer}}]{les06}
\bibinfo{author}{\bibfnamefont{T.}~\bibnamefont{Lesinski}},
  \bibinfo{author}{\bibfnamefont{K.}~\bibnamefont{Bennaceur}},
  \bibinfo{author}{\bibfnamefont{T.}~\bibnamefont{Duguet}}, \bibnamefont{and}
  \bibinfo{author}{\bibfnamefont{J.}~\bibnamefont{Meyer}},
  \bibinfo{journal}{Phys. Rev. C} \textbf{\bibinfo{volume}{74}},
  \bibinfo{pages}{044315} (\bibinfo{year}{2006}).

\bibitem[{\citenamefont{Dobaczewski et~al.}(1984)\citenamefont{Dobaczewski,
  Flocard, and Treiner}}]{dob84}
\bibinfo{author}{\bibfnamefont{J.}~\bibnamefont{Dobaczewski}},
  \bibinfo{author}{\bibfnamefont{H.}~\bibnamefont{Flocard}}, \bibnamefont{and}
  \bibinfo{author}{\bibfnamefont{J.}~\bibnamefont{Treiner}},
  \bibinfo{journal}{Nucl. Phys. A} \textbf{\bibinfo{volume}{422}},
  \bibinfo{pages}{103} (\bibinfo{year}{1984}).

\bibitem[{\citenamefont{Cao et~al.}(2006)\citenamefont{Cao, Lombardo, Shen, and
  Van~Giai}}]{cao06}
\bibinfo{author}{\bibfnamefont{L.}~\bibnamefont{Cao}},
  \bibinfo{author}{\bibfnamefont{U.}~\bibnamefont{Lombardo}},
  \bibinfo{author}{\bibfnamefont{C.}~\bibnamefont{Shen}}, \bibnamefont{and}
  \bibinfo{author}{\bibfnamefont{N.}~\bibnamefont{Van~Giai}},
  \bibinfo{journal}{Phys. Rev. C} \textbf{\bibinfo{volume}{73}},
  \bibinfo{pages}{014313} (\bibinfo{year}{2006}).

\bibitem[{\citenamefont{Hellemans et~al.}(2013)\citenamefont{Hellemans,
  Pastore, Duguet, Bennaceur, Davesne, Meyer, Bender, and Heenen}}]{hel13}
\bibinfo{author}{\bibfnamefont{V.}~\bibnamefont{Hellemans}},
  \bibinfo{author}{\bibfnamefont{A.}~\bibnamefont{Pastore}},
  \bibinfo{author}{\bibfnamefont{T.}~\bibnamefont{Duguet}},
  \bibinfo{author}{\bibfnamefont{K.}~\bibnamefont{Bennaceur}},
  \bibinfo{author}{\bibfnamefont{D.}~\bibnamefont{Davesne}},
  \bibinfo{author}{\bibfnamefont{J.}~\bibnamefont{Meyer}},
  \bibinfo{author}{\bibfnamefont{M.}~\bibnamefont{Bender}}, \bibnamefont{and}
  \bibinfo{author}{\bibfnamefont{P.-H.} \bibnamefont{Heenen}},
  \bibinfo{journal}{Phys. Rev. C} \textbf{\bibinfo{volume}{88}},
  \bibinfo{pages}{064323} (\bibinfo{year}{2013}).

\bibitem[{\citenamefont{Pastore et~al.}(2015)\citenamefont{Pastore, Davesne,
  and Navarro}}]{pas14D}
\bibinfo{author}{\bibfnamefont{A.}~\bibnamefont{Pastore}},
  \bibinfo{author}{\bibfnamefont{D.}~\bibnamefont{Davesne}}, \bibnamefont{and}
  \bibinfo{author}{\bibfnamefont{J.}~\bibnamefont{Navarro}},
  \bibinfo{journal}{Phys. Rep.} \textbf{\bibinfo{volume}{563}},
  \bibinfo{pages}{1} (\bibinfo{year}{2015}), ISSN \bibinfo{issn}{0370-1573}.

\bibitem[{\citenamefont{Pastore et~al.}(2013)\citenamefont{Pastore, Davesne,
  Bennaceur, Meyer, and Hellemans}}]{pas13}
\bibinfo{author}{\bibfnamefont{A.}~\bibnamefont{Pastore}},
  \bibinfo{author}{\bibfnamefont{D.}~\bibnamefont{Davesne}},
  \bibinfo{author}{\bibfnamefont{K.}~\bibnamefont{Bennaceur}},
  \bibinfo{author}{\bibfnamefont{J.}~\bibnamefont{Meyer}}, \bibnamefont{and}
  \bibinfo{author}{\bibfnamefont{V.}~\bibnamefont{Hellemans}},
  \bibinfo{journal}{Phys. Scripta} \textbf{\bibinfo{volume}{2013}},
  \bibinfo{pages}{014014} (\bibinfo{year}{2013}).

\bibitem[{\citenamefont{Fracasso et~al.}(2012)\citenamefont{Fracasso, Suckling,
  and Stevenson}}]{fra12}
\bibinfo{author}{\bibfnamefont{S.}~\bibnamefont{Fracasso}},
  \bibinfo{author}{\bibfnamefont{E.~B.} \bibnamefont{Suckling}},
  \bibnamefont{and} \bibinfo{author}{\bibfnamefont{P.~D.}
  \bibnamefont{Stevenson}}, \bibinfo{journal}{Phys. Rev. C}
  \textbf{\bibinfo{volume}{86}}, \bibinfo{pages}{044303}
  (\bibinfo{year}{2012}).

\bibitem[{\citenamefont{Hellemans et~al.}(2012)\citenamefont{Hellemans, Heenen,
  and Bender}}]{hel12}
\bibinfo{author}{\bibfnamefont{V.}~\bibnamefont{Hellemans}},
  \bibinfo{author}{\bibfnamefont{P.-H.} \bibnamefont{Heenen}},
  \bibnamefont{and} \bibinfo{author}{\bibfnamefont{M.}~\bibnamefont{Bender}},
  \bibinfo{journal}{Phys. Rev. C} \textbf{\bibinfo{volume}{85}},
  \bibinfo{pages}{014326} (\bibinfo{year}{2012}).

\bibitem[{\citenamefont{Perli\ifmmode~\acute{n}\else \'{n}\fi{}ska
  et~al.}(2004)\citenamefont{Perli\ifmmode~\acute{n}\else \'{n}\fi{}ska,
  Rohozi\ifmmode~\acute{n}\else \'{n}\fi{}ski, Dobaczewski, and
  Nazarewicz}}]{per04}
\bibinfo{author}{\bibfnamefont{E.}~\bibnamefont{Perli\ifmmode~\acute{n}\else
  \'{n}\fi{}ska}}, \bibinfo{author}{\bibfnamefont{S.~G.}
  \bibnamefont{Rohozi\ifmmode~\acute{n}\else \'{n}\fi{}ski}},
  \bibinfo{author}{\bibfnamefont{J.}~\bibnamefont{Dobaczewski}},
  \bibnamefont{and}
  \bibinfo{author}{\bibfnamefont{W.}~\bibnamefont{Nazarewicz}},
  \bibinfo{journal}{Phys. Rev. C} \textbf{\bibinfo{volume}{69}},
  \bibinfo{pages}{014316} (\bibinfo{year}{2004}).

\bibitem[{\citenamefont{Toivanen et~al.}(2010)\citenamefont{Toivanen, Carlsson,
  Dobaczewski, Mizuyama, Rodriguez-Guzman, Toivanen, and Vesely}}]{toi10}
\bibinfo{author}{\bibfnamefont{J.}~\bibnamefont{Toivanen}},
  \bibinfo{author}{\bibfnamefont{B.~G.} \bibnamefont{Carlsson}},
  \bibinfo{author}{\bibfnamefont{J.}~\bibnamefont{Dobaczewski}},
  \bibinfo{author}{\bibfnamefont{K.}~\bibnamefont{Mizuyama}},
  \bibinfo{author}{\bibfnamefont{R.~R.} \bibnamefont{Rodriguez-Guzman}},
  \bibinfo{author}{\bibfnamefont{P.}~\bibnamefont{Toivanen}}, \bibnamefont{and}
  \bibinfo{author}{\bibfnamefont{P.}~\bibnamefont{Vesely}},
  \bibinfo{journal}{Phys. Rev. C} \textbf{\bibinfo{volume}{81}},
  \bibinfo{pages}{034312} (\bibinfo{year}{2010}).

\bibitem[{\citenamefont{Carlsson et~al.}(2012)\citenamefont{Carlsson, Toivanen,
  and Pastore}}]{car12}
\bibinfo{author}{\bibfnamefont{B.~G.} \bibnamefont{Carlsson}},
  \bibinfo{author}{\bibfnamefont{J.}~\bibnamefont{Toivanen}}, \bibnamefont{and}
  \bibinfo{author}{\bibfnamefont{A.}~\bibnamefont{Pastore}},
  \bibinfo{journal}{Phys. Rev. C} \textbf{\bibinfo{volume}{86}},
  \bibinfo{pages}{014307} (\bibinfo{year}{2012}).

\bibitem[{\citenamefont{Dobaczewski and Dudek}(1995)}]{dob95}
\bibinfo{author}{\bibfnamefont{J.}~\bibnamefont{Dobaczewski}} \bibnamefont{and}
  \bibinfo{author}{\bibfnamefont{J.}~\bibnamefont{Dudek}},
  \bibinfo{journal}{Phys. Rev. C} \textbf{\bibinfo{volume}{52}},
  \bibinfo{pages}{1827} (\bibinfo{year}{1995}).

\bibitem[{\citenamefont{Lesinski et~al.}(2007)\citenamefont{Lesinski, Bender,
  Bennaceur, Duguet, and Meyer}}]{les07}
\bibinfo{author}{\bibfnamefont{T.}~\bibnamefont{Lesinski}},
  \bibinfo{author}{\bibfnamefont{M.}~\bibnamefont{Bender}},
  \bibinfo{author}{\bibfnamefont{K.}~\bibnamefont{Bennaceur}},
  \bibinfo{author}{\bibfnamefont{T.}~\bibnamefont{Duguet}}, \bibnamefont{and}
  \bibinfo{author}{\bibfnamefont{J.}~\bibnamefont{Meyer}},
  \bibinfo{journal}{Phys. Rev. C} \textbf{\bibinfo{volume}{76}},
  \bibinfo{pages}{014312} (\bibinfo{year}{2007}).

\bibitem[{\citenamefont{Roca-Maza et~al.}(2012)\citenamefont{Roca-Maza, Col\`o,
  and Sagawa}}]{roc12}
\bibinfo{author}{\bibfnamefont{X.}~\bibnamefont{Roca-Maza}},
  \bibinfo{author}{\bibfnamefont{G.}~\bibnamefont{Col\`o}}, \bibnamefont{and}
  \bibinfo{author}{\bibfnamefont{H.}~\bibnamefont{Sagawa}},
  \bibinfo{journal}{Phys. Rev. C} \textbf{\bibinfo{volume}{86}},
  \bibinfo{pages}{031306} (\bibinfo{year}{2012}).

\bibitem[{\citenamefont{Goriely et~al.}(2013)\citenamefont{Goriely, Chamel, and
  Pearson}}]{gor13bsk}
\bibinfo{author}{\bibfnamefont{S.}~\bibnamefont{Goriely}},
  \bibinfo{author}{\bibfnamefont{N.}~\bibnamefont{Chamel}}, \bibnamefont{and}
  \bibinfo{author}{\bibfnamefont{J.~M.} \bibnamefont{Pearson}},
  \bibinfo{journal}{Phys. Rev. C} \textbf{\bibinfo{volume}{88}},
  \bibinfo{pages}{061302} (\bibinfo{year}{2013}).

\bibitem[{\citenamefont{Beiner et~al.}(1975)\citenamefont{Beiner, Flocard,
  Van~Giai, and Quentin}}]{bei75}
\bibinfo{author}{\bibfnamefont{M.}~\bibnamefont{Beiner}},
  \bibinfo{author}{\bibfnamefont{H.}~\bibnamefont{Flocard}},
  \bibinfo{author}{\bibfnamefont{N.}~\bibnamefont{Van~Giai}}, \bibnamefont{and}
  \bibinfo{author}{\bibfnamefont{P.}~\bibnamefont{Quentin}},
  \bibinfo{journal}{Nucl. Phys. A} \textbf{\bibinfo{volume}{238}},
  \bibinfo{pages}{29} (\bibinfo{year}{1975}).

\bibitem[{\citenamefont{Reinhard et~al.}(1999)\citenamefont{Reinhard, Dean,
  Nazarewicz, Dobaczewski, Maruhn, and Strayer}}]{rei99}
\bibinfo{author}{\bibfnamefont{P.-G.} \bibnamefont{Reinhard}},
  \bibinfo{author}{\bibfnamefont{D.}~\bibnamefont{Dean}},
  \bibinfo{author}{\bibfnamefont{W.}~\bibnamefont{Nazarewicz}},
  \bibinfo{author}{\bibfnamefont{J.}~\bibnamefont{Dobaczewski}},
  \bibinfo{author}{\bibfnamefont{J.}~\bibnamefont{Maruhn}}, \bibnamefont{and}
  \bibinfo{author}{\bibfnamefont{M.}~\bibnamefont{Strayer}},
  \bibinfo{journal}{Phys. Rev. C} \textbf{\bibinfo{volume}{60}},
  \bibinfo{pages}{014316} (\bibinfo{year}{1999}).

\bibitem[{\citenamefont{Pastore
  et~al.}(2012{\natexlab{a}})\citenamefont{Pastore, Davesne, Lallouet, Martini,
  Bennaceur, and Meyer}}]{pas12}
\bibinfo{author}{\bibfnamefont{A.}~\bibnamefont{Pastore}},
  \bibinfo{author}{\bibfnamefont{D.}~\bibnamefont{Davesne}},
  \bibinfo{author}{\bibfnamefont{Y.}~\bibnamefont{Lallouet}},
  \bibinfo{author}{\bibfnamefont{M.}~\bibnamefont{Martini}},
  \bibinfo{author}{\bibfnamefont{K.}~\bibnamefont{Bennaceur}},
  \bibnamefont{and} \bibinfo{author}{\bibfnamefont{J.}~\bibnamefont{Meyer}},
  \bibinfo{journal}{Phys. Rev. C} \textbf{\bibinfo{volume}{85}},
  \bibinfo{pages}{054317} (\bibinfo{year}{2012}{\natexlab{a}}).

\bibitem[{\citenamefont{B\"ackman et~al.}(1979)\citenamefont{B\"ackman,
  Sj\"oberg, and Jackson}}]{bac79}
\bibinfo{author}{\bibfnamefont{S.}~\bibnamefont{B\"ackman}},
  \bibinfo{author}{\bibfnamefont{O.}~\bibnamefont{Sj\"oberg}},
  \bibnamefont{and} \bibinfo{author}{\bibfnamefont{A.}~\bibnamefont{Jackson}},
  \bibinfo{journal}{Nucl. Phys. A} \textbf{\bibinfo{volume}{321}},
  \bibinfo{pages}{10} (\bibinfo{year}{1979}).

\bibitem[{\citenamefont{Chamel et~al.}(2009)\citenamefont{Chamel, Goriely, and
  Pearson}}]{cha09}
\bibinfo{author}{\bibfnamefont{N.}~\bibnamefont{Chamel}},
  \bibinfo{author}{\bibfnamefont{S.}~\bibnamefont{Goriely}}, \bibnamefont{and}
  \bibinfo{author}{\bibfnamefont{J.~M.} \bibnamefont{Pearson}},
  \bibinfo{journal}{Phys. Rev. C} \textbf{\bibinfo{volume}{80}},
  \bibinfo{pages}{065804} (\bibinfo{year}{2009}).

\bibitem[{\citenamefont{Terasaki et~al.}(2005)\citenamefont{Terasaki, Engel,
  Bender, Dobaczewski, Nazarewicz, and Stoitsov}}]{ter05}
\bibinfo{author}{\bibfnamefont{J.}~\bibnamefont{Terasaki}},
  \bibinfo{author}{\bibfnamefont{J.}~\bibnamefont{Engel}},
  \bibinfo{author}{\bibfnamefont{M.}~\bibnamefont{Bender}},
  \bibinfo{author}{\bibfnamefont{J.}~\bibnamefont{Dobaczewski}},
  \bibinfo{author}{\bibfnamefont{W.}~\bibnamefont{Nazarewicz}},
  \bibnamefont{and} \bibinfo{author}{\bibfnamefont{M.}~\bibnamefont{Stoitsov}},
  \bibinfo{journal}{Phys. Rev. C} \textbf{\bibinfo{volume}{71}},
  \bibinfo{pages}{034310} (\bibinfo{year}{2005}).

\bibitem[{\citenamefont{Losa et~al.}(2010)\citenamefont{Losa, Pastore,
  D\o{}ssing, Vigezzi, and Broglia}}]{los10}
\bibinfo{author}{\bibfnamefont{C.}~\bibnamefont{Losa}},
  \bibinfo{author}{\bibfnamefont{A.}~\bibnamefont{Pastore}},
  \bibinfo{author}{\bibfnamefont{T.}~\bibnamefont{D\o{}ssing}},
  \bibinfo{author}{\bibfnamefont{E.}~\bibnamefont{Vigezzi}}, \bibnamefont{and}
  \bibinfo{author}{\bibfnamefont{R.~A.} \bibnamefont{Broglia}},
  \bibinfo{journal}{Phys. Rev. C} \textbf{\bibinfo{volume}{81}},
  \bibinfo{pages}{064307} (\bibinfo{year}{2010}).

\bibitem[{\citenamefont{Lechaftois et~al.}(2015)\citenamefont{Lechaftois,
  P\'eru, and Deloncle}}]{lef15}
\bibinfo{author}{\bibfnamefont{F.}~\bibnamefont{Lechaftois}},
  \bibinfo{author}{\bibfnamefont{S.}~\bibnamefont{P\'eru}}, \bibnamefont{and}
  \bibinfo{author}{\bibfnamefont{I.}~\bibnamefont{Deloncle}},
  \bibinfo{journal}{private com.}  (\bibinfo{year}{2015}).

\bibitem[{\citenamefont{Ring and Schuck}(1980)}]{Book:Ring1980}
\bibinfo{author}{\bibfnamefont{P.}~\bibnamefont{Ring}} \bibnamefont{and}
  \bibinfo{author}{\bibfnamefont{P.}~\bibnamefont{Schuck}},
  \emph{\bibinfo{title}{{The Nuclear Many-Body Problem}}}
  (\bibinfo{publisher}{Springer-Verlag}, \bibinfo{year}{1980}).

\bibitem[{\citenamefont{Vesel\'y et~al.}(2012)\citenamefont{Vesel\'y, Toivanen,
  Carlsson, Dobaczewski, Michel, and Pastore}}]{ves12}
\bibinfo{author}{\bibfnamefont{P.}~\bibnamefont{Vesel\'y}},
  \bibinfo{author}{\bibfnamefont{J.}~\bibnamefont{Toivanen}},
  \bibinfo{author}{\bibfnamefont{B.~G.} \bibnamefont{Carlsson}},
  \bibinfo{author}{\bibfnamefont{J.}~\bibnamefont{Dobaczewski}},
  \bibinfo{author}{\bibfnamefont{N.}~\bibnamefont{Michel}}, \bibnamefont{and}
  \bibinfo{author}{\bibfnamefont{A.}~\bibnamefont{Pastore}},
  \bibinfo{journal}{Phys. Rev. C} \textbf{\bibinfo{volume}{86}},
  \bibinfo{pages}{024303} (\bibinfo{year}{2012}).

\bibitem[{\citenamefont{Suzuki et~al.}(1990)\citenamefont{Suzuki, Ikeda, and
  Sato}}]{suz90}
\bibinfo{author}{\bibfnamefont{Y.}~\bibnamefont{Suzuki}},
  \bibinfo{author}{\bibfnamefont{K.}~\bibnamefont{Ikeda}}, \bibnamefont{and}
  \bibinfo{author}{\bibfnamefont{H.}~\bibnamefont{Sato}},
  \bibinfo{journal}{Progr. Theor. Phys.} \textbf{\bibinfo{volume}{83}},
  \bibinfo{pages}{180} (\bibinfo{year}{1990}).

\bibitem[{\citenamefont{Nakatsukasa et~al.}(2007)\citenamefont{Nakatsukasa,
  Inakura, and Yabana}}]{nak07}
\bibinfo{author}{\bibfnamefont{T.}~\bibnamefont{Nakatsukasa}},
  \bibinfo{author}{\bibfnamefont{T.}~\bibnamefont{Inakura}}, \bibnamefont{and}
  \bibinfo{author}{\bibfnamefont{K.}~\bibnamefont{Yabana}},
  \bibinfo{journal}{Phys. Rev. C} \textbf{\bibinfo{volume}{76}},
  \bibinfo{pages}{024318} (\bibinfo{year}{2007}).

\bibitem[{\citenamefont{Kortelainen}(2015)}]{kor15}
\bibinfo{author}{\bibfnamefont{M.}~\bibnamefont{Kortelainen}},
  \bibinfo{journal}{private communication}  (\bibinfo{year}{2015}).

\bibitem[{\citenamefont{Stoitsov et~al.}(2011)\citenamefont{Stoitsov,
  Kortelainen, Nakatsukasa, Losa, and Nazarewicz}}]{sto11}
\bibinfo{author}{\bibfnamefont{M.}~\bibnamefont{Stoitsov}},
  \bibinfo{author}{\bibfnamefont{M.}~\bibnamefont{Kortelainen}},
  \bibinfo{author}{\bibfnamefont{T.}~\bibnamefont{Nakatsukasa}},
  \bibinfo{author}{\bibfnamefont{C.}~\bibnamefont{Losa}}, \bibnamefont{and}
  \bibinfo{author}{\bibfnamefont{W.}~\bibnamefont{Nazarewicz}},
  \bibinfo{journal}{Phys. Rev. C} \textbf{\bibinfo{volume}{84}},
  \bibinfo{pages}{041305} (\bibinfo{year}{2011}).

\bibitem[{\citenamefont{Garcia-Recio et~al.}(1992)\citenamefont{Garcia-Recio,
  Navarro, Giai, and Salcedo}}]{gar92}
\bibinfo{author}{\bibfnamefont{C.}~\bibnamefont{Garcia-Recio}},
  \bibinfo{author}{\bibfnamefont{J.}~\bibnamefont{Navarro}},
  \bibinfo{author}{\bibfnamefont{N.~V.} \bibnamefont{Giai}}, \bibnamefont{and}
  \bibinfo{author}{\bibfnamefont{L.}~\bibnamefont{Salcedo}},
  \bibinfo{journal}{Ann. Phys. (NY)} \textbf{\bibinfo{volume}{214}},
  \bibinfo{pages}{293} (\bibinfo{year}{1992}).

\bibitem[{\citenamefont{Davesne
  et~al.}(2014{\natexlab{b}})\citenamefont{Davesne, Pastore, and
  Navarro}}]{dav14A}
\bibinfo{author}{\bibfnamefont{D.}~\bibnamefont{Davesne}},
  \bibinfo{author}{\bibfnamefont{A.}~\bibnamefont{Pastore}}, \bibnamefont{and}
  \bibinfo{author}{\bibfnamefont{J.}~\bibnamefont{Navarro}},
  \bibinfo{journal}{Phys. Rev. C} \textbf{\bibinfo{volume}{89}},
  \bibinfo{pages}{044302} (\bibinfo{year}{2014}{\natexlab{b}}).

\bibitem[{\citenamefont{Ducoin et~al.}(2008)\citenamefont{Ducoin,
  Provid{\^e}ncia, Santos, Brito, and Chomaz}}]{duc08}
\bibinfo{author}{\bibfnamefont{C.}~\bibnamefont{Ducoin}},
  \bibinfo{author}{\bibfnamefont{C.}~\bibnamefont{Provid{\^e}ncia}},
  \bibinfo{author}{\bibfnamefont{A.~M.} \bibnamefont{Santos}},
  \bibinfo{author}{\bibfnamefont{L.}~\bibnamefont{Brito}}, \bibnamefont{and}
  \bibinfo{author}{\bibfnamefont{P.}~\bibnamefont{Chomaz}},
  \bibinfo{journal}{Phys. Rev. C} \textbf{\bibinfo{volume}{78}},
  \bibinfo{pages}{055801} (\bibinfo{year}{2008}).

\bibitem[{\citenamefont{Nakada}(2010)}]{nak03}
\bibinfo{author}{\bibfnamefont{H.}~\bibnamefont{Nakada}},
  \bibinfo{journal}{Phys. Rev. C} \textbf{\bibinfo{volume}{81}},
  \bibinfo{pages}{027301} (\bibinfo{year}{2010}).

\bibitem[{\citenamefont{Anguiano et~al.}(2012)\citenamefont{Anguiano, Grasso,
  Co', De~Donno, and Lallena}}]{ang12}
\bibinfo{author}{\bibfnamefont{M.}~\bibnamefont{Anguiano}},
  \bibinfo{author}{\bibfnamefont{M.}~\bibnamefont{Grasso}},
  \bibinfo{author}{\bibfnamefont{G.}~\bibnamefont{Co'}},
  \bibinfo{author}{\bibfnamefont{V.}~\bibnamefont{De~Donno}}, \bibnamefont{and}
  \bibinfo{author}{\bibfnamefont{A.~M.} \bibnamefont{Lallena}},
  \bibinfo{journal}{Phys. Rev. C} \textbf{\bibinfo{volume}{86}},
  \bibinfo{pages}{054302} (\bibinfo{year}{2012}).

\bibitem[{\citenamefont{Grasso and Anguiano}(2013)}]{gra13}
\bibinfo{author}{\bibfnamefont{M.}~\bibnamefont{Grasso}} \bibnamefont{and}
  \bibinfo{author}{\bibfnamefont{M.}~\bibnamefont{Anguiano}},
  \bibinfo{journal}{Phys. Rev. C} \textbf{\bibinfo{volume}{88}},
  \bibinfo{pages}{054328} (\bibinfo{year}{2013}).

\bibitem[{\citenamefont{Pastore
  et~al.}(2012{\natexlab{b}})\citenamefont{Pastore, Bennaceur, Davesne, and
  Meyer}}]{pas11}
\bibinfo{author}{\bibfnamefont{A.}~\bibnamefont{Pastore}},
  \bibinfo{author}{\bibfnamefont{K.}~\bibnamefont{Bennaceur}},
  \bibinfo{author}{\bibfnamefont{D.}~\bibnamefont{Davesne}}, \bibnamefont{and}
  \bibinfo{author}{\bibfnamefont{J.}~\bibnamefont{Meyer}},
  \bibinfo{journal}{Int.J.Mod.Phys.} \textbf{\bibinfo{volume}{E21}},
  \bibinfo{pages}{1250040} (\bibinfo{year}{2012}{\natexlab{b}}).

\end{thebibliography}

\end{document}